\documentclass[lettersize,journal]{IEEEtran}
\usepackage{amsmath,amssymb,amsfonts}
\usepackage{algorithmic}
\usepackage{algorithm}
\usepackage{array}
\usepackage{subfig}
\usepackage{textcomp}
\usepackage{stfloats}
\usepackage{url}
\usepackage{verbatim}
\usepackage{graphicx}
\usepackage{cite}
\usepackage{color}
\usepackage{epstopdf}
\usepackage{makecell}
\usepackage{amsthm}
\DeclareMathAlphabet\mathbfcal{OMS}{cmsy}{b}{n}
\newcommand{\Rmnum}[1]{\uppercase\expandafter{\romannumeral #1}}

\begin{document}

\title{Matched Filtering-Based Channel Estimation for AFDM Systems in Doubly Selective Channels}

\author{Xiangjun Li, Zilong Liu, Zhengchun Zhou, and Pingzhi Fan
        % <-this % stops a space
%\thanks{This paper was produced by the IEEE Publication Technology Group. They are in Piscataway, NJ.}% <-this % stops a space
\thanks{Xiangjun Li, Zhengchun Zhou and Pingzhi Fan are with the School of Info Sci \& Tech, Southwest Jiaotong University, Chengdu, China. 
Zilong Liu is with the School of Computer Science and Electronics Engineering, University of Essex, U. K.
%The work of Pingzhi Fan and Xiangjun Li was supported by NSFC project No.62020106001 and No.U23A20274, the work of Qianli Wang was supported by NSFC No.62301455 and No.62350610267, the work of Zilong Liu was supported by UK EPSRC under Grants EP/X035352/1 and EP/Y000986/1.
Corresponding author: Zhengchun Zhou. Emails: lxj@my.swjtu.edu.cn; zilong.liu@essex.ac.uk; zzc@swjtu.edu.cn; pzfan@swjtu.edu.cn.
}
}
%\IEEEpubid{\begin{minipage}{\textwidth}\ \centering
%		Copyright \copyright 20xx IEEE. Personal use of this material is permitted. \\
%		However, permission to use this material for any other purposes must be obtained from the IEEE by sending a request to pubs-permissions@ieee.org.
%\end{minipage}}

\maketitle

\begin{abstract}
Affine frequency division multiplexing (AFDM) has recently emerged as an excellent backward-compatible 6G waveform. In this paper, we study matched filtering (MF) assisted channel estimation (CE) for AFDM systems in complex doubly selective channels. By deriving the complete input-output relationship of the continuous-time signal, the inter-chirp-carrier interference, signal-to-interference-plus-noise ratio (SINR), and the effective SINR loss of AFDM, are investigated in discrete affine Fourier transform (DAFT) domain. Further, we propose two low-complexity methods for constructing the channel matrix by taking advantage of its inherent discrete Fourier transform structure and the staircase structure of the piecewise functions in the channel matrix, respectively. It is shown that complexity reduction by at least two orders of magnitude can be achieved for a large number of chirp subcarriers.
For the CE problem in doubly selective channels, we introduce an MF assisted CE scheme.
This allows us to sequentially estimate the parameters of each path by exploiting the separability and approximate orthogonality of different paths in the DAFT domain, thus leading to significantly reduced complexity. Furthermore, based on generalized Fibonacci search (GFS), an MF-GFS scheme is proposed to avoid significantly redundant computation, which can be extended to typical wide-band systems. Extensive simulation results indicate that the proposed schemes offer superior advantages in terms of their improved communication performance and lower complexity.

\end{abstract}


\section{Introduction}
%\IEEEPARstart{D}{oppler} shift in fast time-varying channels significantly deteriorates the performance of orthogonal frequency division multiplexing. In contrast, the orthogonal time-frequency space (OTFS) modulation has garnered increasing attention in high-mobility wireless communication scenarios due to its robustness in representing doubly fractional  channels \cite{hadani1,P. Raviteja_1,P. Raviteja_2}.

\IEEEPARstart{T}{he} sixth generation (6G) communication systems are deemed to support ultra-reliable, low-latency, and high-rate communications in highly dynamic scenarios, such as vehicle-to-everything (V2X) systems, unmanned aerial vehicles, high-speed trains, and low-earth-orbit (LEO) satellites. Traditional orthogonal frequency-division multiplexing (OFDM) may be infeasible due to significant inter-carrier interference caused by high mobility \cite{T. Wang, L. Rugini, Liu-V2X-2022}.

Several waveforms that can adapt to high-mobility scenarios have been studied.
Among many others, a representative waveform is orthogonal time-frequency space (OTFS) whose information symbols are transmitted in the delay-Doppler (DD) domain through two-dimensional (2D) orthogonal basis functions \cite{R. Hadani_1, Z. Wei_1, P. Raviteja_1}. Since each information symbol in the DD domain spans the entire time-frequency grid, OTFS is able to achieve a significant improvement in error rate performance compared to OFDM \cite{Z. Wei_1,P. Raviteja_1,S. Li,P. Raviteja_2,Z. Wei_2,Q. Wang,X. Li}.  In addition, Zak-OTFS \cite{S. Gopalam,F. Lampel} has emerged as a strong competitor to conventional OTFS in recent years, offering an alternative Zak-transform-based formulation that preserves the DD domain advantages while enabling flexible waveform implementations.
However, OTFS requires radical change for the transceiver design and hence may not permit a seamless integration into the legacy OFDM based wireless systems. 

Recently, affine frequency division multiplexing (AFDM) has emerged as an excellent backward-compatible 6G waveform for efficient and reliable high-mobility communications \cite{Bemani_1, Bemani_2, Bemani_3,H. Yin,Z. Sui,Hyeon,Q. Li,Bemani_5}. With minimum modification of OFDM, AFDM modulates the data symbols using multiple orthogonal chirp-carriers. The  modulation is carried out through discrete affine Fourier transform (DAFT), 
%which is a generalization of the discrete Fourier transform (DFT) and the discrete Fresnel transform
enabling efficient mapping between the DAFT domain and the time domain. By appropriately tuning the chirp rate according to the Doppler profile of the channel, AFDM enables proper spreading in the time-frequency domain, thus allowing it to achieve the full diversity over the doubly selective channels. Besides, the sparsity and compactness of the channel in the DAFT domain can be exploited for reducing the pilot overhead. Special cases of AFDM include DAFT-OFDM \cite{T. Erseghe} and orthogonal chirp division multiplexing (OCDM) \cite{X. Ouyang}, yet they may not be able to achieve the full diversity \cite{Bemani_3}.

A plethora of recent works have further explored AFDM. 
By flexibly selecting the chirp parameter, AFDM can achieve significant performance improvements in several aspects.
Inspired by the index modulation (IM) for OFDM \cite{E. Başar}, IM-AFDM systems were studied in \cite{Y. Tao, J. Zhu_1, G. Liu, H. S. Rou,Hyeon_1} for improving the spectrum efficiency.
From the PAPR reduction aspect, the chirp-permuted AFDM was employed in \cite{Hyeon_1}, while the grouped pre-chirp selection algorithm was proposed in \cite{H. Yuan}.

%Considering that the DAFT parameter $c_2$ does not affect the orthogonality between chirp carriers, a new class of IM-AFDM was investigated in \cite{G. Liu, H. S. Rou} by leveraging the activation state of $c_2$.
The integration of generalized spatial modulation (GSM) and AFDM, called GSM-AFDM, was studied in \cite{Sui2025} to design low-power and high-performance multiple-input and multiple-output (MIMO) systems. For 6G integrated sensing and communications (ISAC), several studies demonstrated the advantages of AFDM empowered ISAC \cite{Y. Ni, A. Bemani_4, J. Zhu, H.Yin_2025} systems. To support massive machine-type high-mobility communications, AFDM was also exploited as the building waveform for sparse code multiple access (SCMA) systems in \cite{Luo2024}. 

Most existing works assume that the receiver has perfect channel information to achieve excellent transmission performance. In practice, however, channel estimation (CE) is generally required for coherent receiving systems.
%\footnote{In this work, we use PE as an inclusive terminology that covers both channel estimation and various sensing tasks in AFDM systems.}
Similar to the traditional CE scheme in OTFS \cite{P. Raviteja_2}, the least square (LS) estimator was investigated in \cite{H. Yin_1} along with a reasonable threshold to estimate the path parameters. By exploiting the sparsity of the channel, the authors in \cite{Benzine_1,Benzine_2,F. Yang} modeled CE as a sparse signal recovery problem and estimated the channel using the compressed sensing (CS) algorithm.

The aforementioned schemes mostly assume that the normalized delay and Doppler values of the paths are integers. Nevertheless, the real-life channel response of each path typically does not align with the grid point in the DAFT domain. That is, the normalized delay and Doppler values may have fractional components, potentially resulting in loss of channel sparsity, as well as degradation of CE performance in the DAFT domain. To address this issue, a CE scheme for MIMO-AFDM based on the diagonal reconstruction of the subchannel matrix was proposed in \cite{H. Yin_2}. Their scheme directly estimates the effective channel matrix instead of the specific channel parameters. An approximate maximum likelihood CE scheme for AFDM was proposed in \cite{Bemani_3} by assuming that the delay of each propagation path is different and the number of paths is known \textit{a priori}. In addition, the joint estimation of all paths in \cite{Bemani_3} results in a high computational complexity.

Based on the correlation between AFDM basis functions and the received signal, this work investigates matched filtering (MF) assisted CE in doubly selective channels for AFDM systems. We differentiate integer-delay-fractional-Doppler (IDFD) and fractional-delay-fractional-Doppler (FDFD) channels and develop advanced MF-CE schemes. %Specifically, for integer-delay-fractional-Doppler (IDFD) channels, a suitable observation model is established to obtain a joint maximum likelihood estimation (JMLE). For lower computational complexity, we exploit the separability and approximate orthogonality of different paths in the DAFT domain. Further, based on the generalized Fibonacci search (GFS) algorithm \cite{Avriel,Subasi,Chong}, the MF-GFS scheme is proposed to further reduce redundant calculations. Additionally, for fractional-delay-Doppler (FDD) channels, a new pilot pattern and another MF-GFS scheme are proposed to address the path ambiguity problem that causes PE performance degradation.
The main contributions are summarized as follows:
\begin{itemize}
	\item{
%		By establishing the complete input-output (I/O) relationship, it is shown that the classical AFDM \cite{Bemani_1,Bemani_2,Bemani_3} can be considered as a special case in our enhanced AFDM when the DD coupling phase is omitted.
%		Firstly, by establishing the complete input-output (I/O) relationship, we show that the DD coupling phase was overlooked in classical AFDM \cite{Bemani_3}. Taking into account of such a DD coupling phase, 
		Firstly, we adopt the input-output (I/O) relationship in the time domain, in alignment with that employed in existing OTFS-related research.
		We derive the corresponding I/O relationship for continuous-time signals and then present a comprehensive analysis on the inter-chirp-carrier-interference (ICCI), signal-to-interference-plus-noise ratio (SINR) and effective SINR loss.
%		To establish the observation model for CE, we derive the complete input-output (I/O) relationship for AFDM systems that is more consistent with the actual physical channel.
%%		We consider the generic channel response in the DD domain and derive the complete input-output (I/O) relationship of AFDM systems. 
%%		Our study reveals that an important DD coupling phase in the channel matrix expression is missing in the existing research works.
%		Our study reveals that the derived I/O relationship is a more general result compared to existing studies which miss an important DD coupling phase.
%		Building upon the derived I/O relationship, we present a comprehensive analysis on the inter-chirp-carrier-interference (ICCI), signal-to-interference-plus-noise ratio (SINR) and effective SINR loss.
%	under integer-delay-Doppler (IDD) channels, IDFD channels and FDFD channels.
	}
	
	\item{
	Secondly, for FDFD channels, we propose two low-complexity methods for fast channel matrix construction. The first method exploits the inherent DFT structure of the channel matrix, enabling efficient computation via  fast fourier transform (FFT). The second method leverages the staircase structure of the piecewise functions in the channel matrix to directly calculate partial sums. For large subcarrier numbers, the proposed methods can reduce the complexity by at least two orders of magnitude.
}

	\item{
		Thirdly, for FDFD channels, we propose an MF CE scheme with joint DD estimation (MF-JE). 
		By leveraging the separability and orthogonality of different paths, our proposed MF scheme is able to eliminate the matrix inversion operation, decouple multipath estimation, and narrow the search range, thus leading to significantly reduced complexity. In addition, by decoupling the DD estimation to further narrow the search region, a more efficient MF (MF-DE) CE scheme is developed at the cost of a slight performance loss.
	}
	
	\item{Fourthly, based on the generalized Fibonacci search (GFS) algorithm\footnote{GFS is an unconstrained nonlinear optimization method for unimodal functions. The ratio of two consecutive generalized Fibonacci numbers (GFNs) approximates the golden ratio. In GFS, two consecutive GFNs are used to non-uniformly divide the search interval, thereby efficiently narrowing the search range, reducing complexity, and enabling more accurate estimation.} \cite{Avriel,Subasi,Chong}, we propose an MF-GFS-DE CE scheme for FDFD channels. It is found that our proposed scheme can reduce the amount of search when estimating fractional parameters, yielding a significant reduction of redundant computations whilst outperforming the proposed MF scheme.} %With the same computational complexity, the estimation performance of the MF-GFS scheme outperforms that of the MF scheme. }
	
	\item{Finally, we extend the proposed MF CE to the IDFD channels in typical wide-band systems (i.e., millimeter-wave communication systems) whereby the bandwidth is sufficiently large to approximate the normalized delay shifts to be integers. 
		Given that the objective function for estimating fractional Doppler can also be demonstrated to be unimodal, an MF-GFS scheme is proposed for IDFD channels. Simulation results indicate that the proposed schemes offer advantages in both computational complexity and performance.}

\end{itemize}

The rest of this paper is organized as follows. Section II reviews the basic concepts of AFDM. Section III analyzes channel in the DAFT domain. Section IV introduces the MF CE scheme and MF-GFS CE scheme. Section V gives the simulation results. Finally, Section VI concludes this paper.

\textit{Notations}: The $m$-th element of vector $\boldsymbol{x}$ is denoted by $x\left[ m\right] $, the element in the $m$-th row and $n$-th column of matrix $\boldsymbol{X}$ is $X\left[ m,n \right] $. $\left( \cdot \right) ^*$, $\left( \cdot \right) ^\mathrm{T}$ and $\left( \cdot \right) ^\mathrm{H}$ denote the conjugate, transpose, and Hermitian operations, respectively. $\mathrm{diag}\left( \cdot \right) $ is to convert a vector into a diagonal matrix or extract the diagonal elements of a matrix. $\lceil \cdot \rceil $, $\lfloor \cdot \rfloor $, and $\lfloor \cdot \rceil $ denote the ceiling, floor, and nearest-integer operators, respectively.
$E\left\{ \cdot \right\} $ is the expected operation.
$\left\{\boldsymbol{x}\times \boldsymbol{y}\right\}$ is the Cartesian product.
$\delta \left( \cdot \right) $ is the Dirac delta function. $\boldsymbol{I}_N$ is the  $N \times N$ identity matrix.
%$\boldsymbol{x}\sim \mathcal{C} \mathcal{N} \left( \boldsymbol{\mu },\boldsymbol{\varSigma } \right) $ indicates that vector $\boldsymbol{x}$ follows a complex Gaussian distribution with mean $\boldsymbol{\mu }$ and variance $\boldsymbol{\varSigma }$. 
$\left| \cdot \right|$ denotes the modulus operation. $\left[ \cdot \right] _N$ is the mod-$N$ operation.

\section{System Models}
%In this section, we will introduce the basic AFDM system model.
\subsection{AFDM modulation}
\begin{figure}[t]
	\centerline{\includegraphics[width=\columnwidth ]{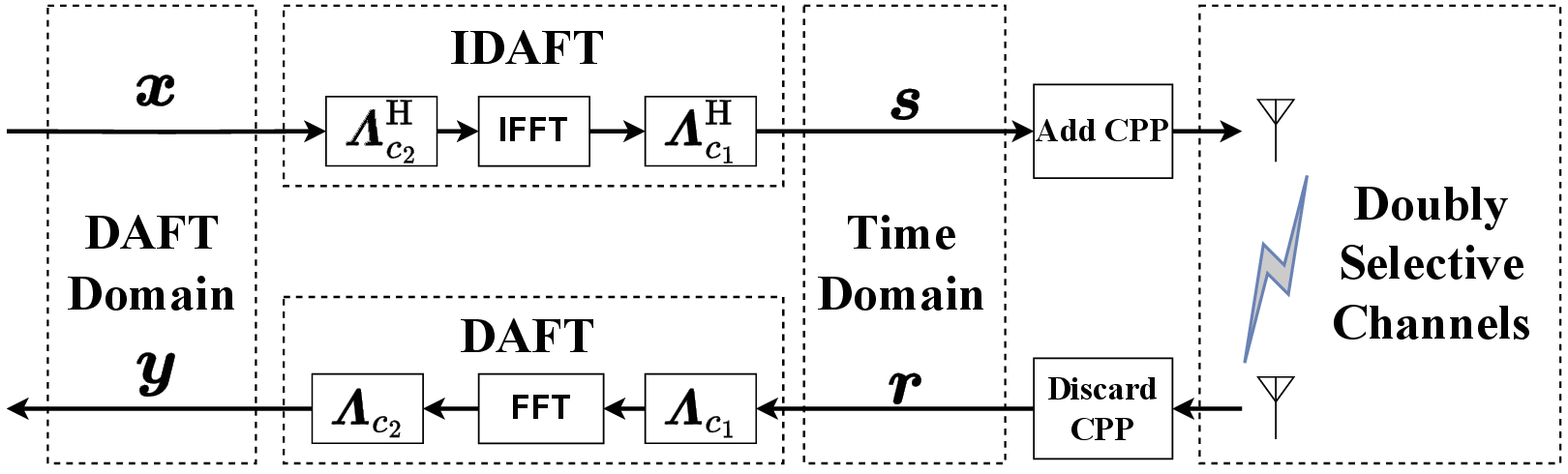}}
	\caption{AFDM system block diagram.}
	\label{Block_Diagram}
\end{figure}
Assume that the number of chirp subcarriers is $N$, the subcarrier spacing is $\varDelta f$ (Hz), the bandwidth is $B=N\varDelta f$, and the duration of an AFDM symbol is $T$.
The DAFT domain axis is sampled at multiple times of $\varDelta f$ to obtain a discrete grid, i.e., $\boldsymbol{\varXi } =\left\{ n\varDelta f,n=0,\cdots ,N-1 \right\} $.
In this paper, the AFDM system is illustrated in Fig. \ref{Block_Diagram}. 
For the transmitter (TX) side, consider a set of $N$ information symbols $\left\{ x\left[ n \right] ,n=0,\cdots ,N-1 \right\} $ from a modulation alphabet $\mathbb{A}$ of size $M_{\mathrm{mod}}$ (e.g. QAM symbols), which are arranged on the DAFT domain grid $\boldsymbol{\varXi }$. 
The AFDM modulator first employs inverse DAFT (IDAFT) to map $x\left[ n \right]$ to $s\left[ \bar{n} \right] $, as shown below:
\begin{equation}\label{AFDM modulation}
	s\left[ \bar{n} \right] =\sum_{n=0}^{N-1}{x\left[ n \right]}\phi _n\left[ \bar{n} \right] ,
\end{equation}
where $\phi _n\left[ \bar{n} \right] =\frac{1}{\sqrt{N}}e^{j2\pi \left( c_1\bar{n}^2+c_2n^2+\frac{\bar{n}n}{N} \right)}$, $\bar{n}\in \left\{ 0,\cdots ,N-1 \right\} $ is the time index, $n\in \left\{ 0,\cdots ,N-1 \right\} $ is the chirp subcarrier index in the DAFT domain, $c_1$ and $c_2$ are DAFT parameters. Note that with a proper $c_1$, the paths are separable in the DAFT domain, enabling AFDM to achieve full diversity in doubly selective channels, i.e., \cite{Bemani_3}
\begin{equation}\label{c1}
	c_1=\frac{2\left( k_{\max}+\xi \right) +1}{2N},
\end{equation}
where $\xi $ is a small non-negative integer introduced to suppress fractional Doppler. This paper sets $c_1$ to the value in \eqref{c1} by default. Other values of $c_1$ may be possible as long as AFDM achieves full diversity.
%$c_2$ can be set to any irrational number or a rational number much smaller than $\small{\frac{1}{2N}}$, as it does not affect the orthogonality between the chirp carriers \cite{Bemani_3}.
\eqref{AFDM modulation} can be rewritten as
\begin{equation}
	\boldsymbol{s}=\boldsymbol{A}^{\mathrm{H}}\boldsymbol{x},
\end{equation}
where $\boldsymbol{x}=\left[ x\left[ 0 \right] ,\cdots ,x\left[ N-1 \right] \right] ^\mathrm{T}\in \mathbb{C} ^{N\times 1}$, $\boldsymbol{s}=\left[ s\left[ 0 \right] ,\cdots ,s\left[ N-1 \right] \right] ^\mathrm{T}\in \mathbb{C} ^{N\times 1}$, $\boldsymbol{A}=\boldsymbol{\varLambda }_{c_2}\boldsymbol{F\varLambda }_{c_1}$ is the normalized DAFT matrix, $\boldsymbol{F}$ is the normalized DFT matrix, $\boldsymbol{\varLambda }_c=\mathrm{diag}\mathrm{(}e^{-j2\pi cn^2},n=0,1,...,N-1)$.
Before transmission, a chirp-periodic prefix (CPP) \cite{Bemani_3} of length $L_\mathrm{CPP}$ is also employed.
To reflect the spectral wrapping inherent in the AFDM signal, the continuous transmit signal in $0\leqslant t<T$ should be expressed as \cite{Bemani_5}
\begin{equation} \label{s_t}
	s\left( t \right) =\frac{1}{\sqrt{N}}\sum_{n=0}^{N-1}{x\left[ n \right] g_{\mathrm{tx}}\left( t \right) \phi _n\left( t \right) ,}
\end{equation}
where $g_{\mathrm{tx}}\left( t \right)$ is the transmit pulse-shaping filter, $\phi _n\left( t \right) =e^{j2\pi \left( c_2n^2+\frac{c_1}{T_{s}^{2}}t^2+\frac{n}{T}t+\varphi _n\left( t \right) \right)}$, $T_s=\frac{T}{N}$ is the sampling period, $\varphi _n\left( t \right) =\sum_{q=0}^{2Nc_1}{\alpha _{n,q}\left( t \right)}$, $q\in \left\{ 0,1,\cdots ,2Nc_1 \right\} $, $\alpha _{n,q}\left( t \right)$ is given by
\begin{equation}
	\alpha _{n,q}\left( t \right) =\begin{cases}	\lfloor \frac{q\left( q-\frac{n}{N} \right)}{2c_1} \rfloor -\frac{q}{T_s}t,&		t_{n,q}\le t<t_{n,q+1}\\	0,&		\mathrm{otherwise}\\\end{cases},
\end{equation}
\begin{equation}
	t_{n,q}=\begin{cases}	0,&		q=0\\	\frac{qN-n}{2Nc_1}T_s,&		1\le q\le 2Nc_1\\\end{cases}.
\end{equation}

%Before transmission, a chirp-periodic prefix (CPP) of length $L_\mathrm{CPP}$ is also employed, serving a similar purpose to the cyclic prefix (CP) used in OFDM systems to mitigate multipath interference, i.e.,
%\begin{equation}
%	s\left( \bar{n} \right) =s[N+\bar{n}]e^{-j2\pi c_1(N^2+2N\bar{n})},\bar{n}=-L_{\mathrm{CPP}},\cdots ,-1,
%\end{equation}
%where $L_\mathrm{CPP}$ is an integer greater than or equal to the maximum normalized delay of the path. When $2Nc_1$ is an integer and $N$ is an even number, CPP is simplified to a CP.

\subsection{Doubly Selective Channels}
%Since AFDM can obtain a full delay-Doppler representation of the doubly selective channels, this paper considers the DD domain channel response to represent the doubly selective channels.
%There are usually only a few scatterers with delay and Doppler in the propagation environment, which means that only a few parameters are needed to model the channel in the DD domain.
Considering the sparsity of the channel, the channel response $h\left( \tau ,\nu \right) $ can be expressed as
\begin{equation}\label{DD channel}
	h\left( \tau ,\nu \right) =\sum_{i=1}^P{h_i}\delta \left( \nu -\nu _i \right) \delta \left( \tau -\tau _i \right),
\end{equation}
where $h_i$, $\tau _i\in \left[ 0,\tau _{\max} \right] $ and $\nu _i\in \left[ -\nu _{\max},\nu _{\max} \right] $ are the channel gain, delay and Doppler parameters of the $i$-th path, respectively, $P$ is the number of paths. 
$\nu _{\max}$ and $\tau _{\max}$ are the maximum Doppler and delay, respectively.
 The normalized delay $l_{\tau _i}$ and Doppler $k_{\nu _i}$ satisfy
\begin{equation}\label{l_k}
	\tau _i=\small{\frac{l_{\tau _i}}{N\varDelta f}},\ \nu _i=k_{\nu _i}\varDelta f,
\end{equation}
where $k_{\nu _i}=\bar{k}_{\nu _i}+\kappa _{\nu _i}\in \left[ -k_{\max},k_{\max} \right] $, $l_{\tau _i}=\bar{l}_{\tau _i}+\iota _{\tau _i}\in \left[ 0,l_{\max} \right] $, $\bar{k}_{\nu _i}$ and $\bar{l}_{\tau _i}$ are both integer parts, $\kappa _{\nu _i}\in \left[ -0.5,0.5 \right] $ and $\iota _{\tau _i}\in \left[ -0.5,0.5 \right] $ are both fractional parts, $k_{\max}=\lceil \frac{N\nu _{\max}}{B} \rceil $, $l_{\max}=\lceil \tau _{\max}B \rceil $. 
%Note that \eqref{DD channel} is a generalized model, allowing for the possibility that each normalized delay may have different normalized Doppler, and vice versa.  
%In addition, when the bandwidth is $N\varDelta f$ and the time-width is $T$, the delay and Doppler resolutions are $\frac{1}{N\varDelta f}$ and $\varDelta f$, respectively.
%Due to the finite $B$ and $T$, the delay and Doppler sampling grids are usually not aligned with the true normalized delay and Doppler values, resulting in fractional delay and Doppler.
Due to the finite $B$ and $T$, there is generally fractional delay and Doppler.
$\nu _{\max}$ and $\tau _{\max}$ can usually be considered known a priori \cite{Z. Wei_2}.
%Since $\nu _{\max}$ and $\tau _{\max}$ can typically be obtained through long-term measurements in practical scenarios \cite{Z. Wei_2}, they are assumed known a priori.
%In this paper, it is assumed that $\tau _{\max}$ and $\nu _{\max}$ are known a priori.
%In practical scenarios, $\nu _{\max}$ and $\tau _{\max}$ can typically be obtained through long-term measurements \cite{Z. Wei_2}. 
%\begin{equation}
%	\nu _{\max}=\underset{\nu _i}{\max}\left\{ \left| \nu _i \right| \right\},\\
%	\tau _{\max}=\underset{\tau _i}{\max}\left\{ \tau _i \right\}.
%\end{equation}
However, the number of scatterers in the environment is usually unpredictable, resulting in the uncertainty of $P$.

After digital-to-analog conversion, the transmitted signal is represented by $s\left( t \right) $. In \cite{Bemani_3}, the received signal $r\left( t \right) $ is
\begin{equation}\label{rreceive0}
	r(t)=\iint{h(\tau ,\nu )s(t-\tau )e^{j2\pi \nu t}\mathrm{d}\tau \mathrm{d}\nu}+\tilde{w}\left( t \right) .
\end{equation}
However, to be consistent with existing OTFS-related research, $r\left( t \right) $ in this paper is given by \cite{R. Hadani_1}
\begin{equation}\label{rreceive}
	r(t)=\iint{h(\tau ,\nu )s(t-\tau )e^{j2\pi \nu \left( t-\tau \right)}\mathrm{d}\tau \mathrm{d}\nu}+\tilde{w}\left( t \right) ,
\end{equation}
where $\tilde{w}\left( t \right) $ is the additive white Gaussian noise.
%\eqref{rreceive} commonly appears in existing OTFS-related research.
The two interpretations of the channel impulse response differ by a term $e^{-j2\pi\nu\tau}$. As long as the notation is consistent, equivalent results can be obtained by using either definition.
%Although these two I/O relationships are different in phase, they are equivalent.

\theoremstyle{remark}
\newtheorem{remark}{\bf Remark}
%\begin{remark}
%	Compared to the existing studies on AFDM that use $e^{j2\pi \nu t}$ to represent phase variations \cite{Bemani_3}, 
%	$e^{j2\pi \nu \left( t-\tau \right)}$ more precisely characterizes the joint effects of delay and Doppler on the signal phase, thereby providing a more accurate reflection of the true physical channel.
%\end{remark}
After discarding CPP, $r\left( t \right) $ is given by
\begin{equation}\label{r_t}
	\begin{aligned}
		r(t)&=\sum_{i=1}^P{h_is\left( t-\tau _i \right) e^{j2\pi \nu _i\left( t-\tau _i \right)}}+\tilde{w}\left( \bar{n} \right) ,
	\end{aligned}
\end{equation}
where $\tilde{w}\left( \bar{n} \right) $ is a zero-mean Gaussian noise with variance $\sigma^2$.
%\eqref{r_n} can be expressed in matrix form as
%\begin{equation}
%	\boldsymbol{r}=\boldsymbol{H}_{\mathrm{T}}\boldsymbol{s}+\tilde{\boldsymbol{w}},
%\end{equation}
%%which can be expressed as
%%\begin{equation}
%%	\boldsymbol{H}_{\mathrm{T}}=\sum_{i=1}^P{h_i}\boldsymbol{\varGamma }_{\mathrm{CPP}_i}\boldsymbol{\varPi }^{l_{\tau _i}}\boldsymbol{\varDelta }^{k_{\nu _i}},
%%\end{equation}
%where $\boldsymbol{H}_{\mathrm{T}}=\sum_{i=1}^P{h_i}\boldsymbol{\varGamma }_{\mathrm{CPP}_i}\boldsymbol{\varPi }^{l_{\tau _i}}\boldsymbol{\varDelta }^{k_{\nu _i}}\in \mathbb{C} ^{N\times N}$ is the channel matrix in the time domain, $\boldsymbol{\varDelta }=\mathrm{diag}\left( z^0,z^1,\cdots z^{N-1} \right)\in \mathbb{C} ^{N\times N}$, $z=e^{\frac{j2\pi}{N}}$,
%$\boldsymbol{\varPi }\in \mathbb{C} ^{N\times N}$ is the forward cyclic-shift matrix,
%$\boldsymbol{\varGamma }_{\mathrm{CPP}_i}\in \mathbb{C} ^{N\times N}$ is used to model CPP \cite{Bemani_3}.
%The delay and Doppler values are modeled by $\boldsymbol{\varPi }$ and $\boldsymbol{\varDelta }$, respectively.
%%For each path, the transmitted signal vector $\boldsymbol{s}$ is modulated by a carrier with frequency $k_{\nu _i}$ and simultaneously undergoes an $l_{\tau _i}$-order cyclic shift. This can be modeled as $\boldsymbol{\varDelta }^{k_{\nu _i}}$ and $\boldsymbol{\varPi }^{l_{\tau _i}}$, respectively.

\subsection{AFDM Demodulation}
At the receiver side, demodulation is performed to obtain $y\left[ m \right] $ in the DAFT domain, i.e.,
\begin{equation}\label{y_m}
	y\left[ m \right] =\sum_{\bar{n}=0}^{N-1}{\left( r\left( t \right) g_{\mathrm{rx}}\left( t \right) \phi _{m}^{*}\left( t \right) \right) \left| _{t=\bar{n}T_S} \right. ,}
\end{equation}
where $g_{\mathrm{rx}}\left( t \right)$ is the receive pulse-shaping filter. For simplicity, this paper adopts rectangular pulse shaping.
%In this paper, $g_{\mathrm{tx}}\left( t \right)$ and $g_{\mathrm{rx}}\left( t \right)$ are rectangular pulses with amplitude $\frac{1}{\sqrt{T}}$ for $0\leqslant t<T$.

%\eqref{y_m} can be rewritten as
%\begin{equation}\label{y_DAFT}
%	\boldsymbol{y}=\boldsymbol{Ar}=\boldsymbol{H}_{\mathrm{eff}}\boldsymbol{x}+\boldsymbol{w},
%\end{equation}
%where $\boldsymbol{y}=\left[ y\left( 0 \right),\cdots ,y\left( N-1 \right) \right] ^{\mathrm{T}}\in \mathbb{C} ^{N\times 1}$, $\boldsymbol{H}_{\mathrm{eff}}$ is the effective channel matrix in the DAFT domain.

\section{Channel Analysis in The DAFT Domain}
This section analyzes the I/O relationship, ICCI, SINR, and the effective SINR loss. It also presents two low-complexity methods for channel matrix construction.
\subsection{I/O Relation of AFDM}
%Since the channel modeling is defined as $\boldsymbol{\varDelta }^{k_{\nu _i}}\boldsymbol{\varPi }^{l_{\tau _i}}$ instead of $\boldsymbol{\varPi }^{l_{\tau _i}}\boldsymbol{\varDelta }^{k_{\nu _i}}$ in the existing work, the derived effective channel matrix in the DAFT domain will lack a phase term with delay and Doppler coupling. 
%This makes the modeling error between the channel model and the real channel larger, leading to significant degradation in system performance. 

\newtheorem{theorem}{\bf Theorem}
\begin{theorem}\label{thm1}
	The $y\left[ m \right] $ can be expressed as
	\begin{equation}\label{y_m_2}
		\begin{aligned}
			y\left[m\right]=\sum_{n=0}^{N-1}{x\left[ n \right] H_\mathrm{eff}\left[ m,n \right]}+w\left[ n \right] ,
		\end{aligned}
	\end{equation}
	\begin{equation}\label{h_w_1}
		\begin{aligned}
			H_\mathrm{eff}\left[ m,n \right] =\sum_{i=1}^P{h_i\mathcal{G} (m,n,k_{\nu _i},l_{\tau _i})},
		\end{aligned}
	\end{equation}
	\begin{equation}\label{G}
		\mathcal{G} (m,n,k_{\nu _i},l_{\tau _i})=\alpha (m,n,k_{\nu _i},l_{\tau _i})\mathcal{F} (m,n,k_{\nu _i},l_{\tau _i}),
	\end{equation}
	\begin{equation}\label{alpha}
		\alpha \left( m,n,k_{\nu _i},l_{\tau _i} \right) =e^{\small{\frac{-j2\pi}{N}\left( -Nc_1l_{\tau _i}^{2}+\left( n+k_{\nu _i} \right) l_{\tau _i}+Nc_2\left( m^2-n^2 \right) \right)}},
	\end{equation}
	\begin{equation}\label{F_2}
		\begin{aligned}
			&\mathcal{F} \left( m,n,l_{\tau _i},k_{\nu _i} \right) \\&=\frac{1}{N}\sum_{\bar{n}=0}^{N-1}{e^{\small{\frac{-j2\pi}{N}}\left( \left( m-n+2Nc_1l_{\tau _i}-k_{\nu _i} \right) \bar{n}-Nd_{\bar{n},n}\iota _{\tau _i} \right)}}\\&=\begin{cases}	\frac{1}{N}\delta \left( \left[ m-n+2Nc_1l_{\tau _i}-k_{\nu _i} \right] _N \right)&		\kappa _{\nu _i}=0,\iota _{\tau _i}=0\\	\frac{1}{N}\frac{e^{\small{-j2\pi}\left( m-n+2Nc_1l_{\tau _i}-k_{\nu _i} \right)}-1}{e^{\small{\frac{-j2\pi}{N}}\left( m-n+2Nc_1l_{\tau _i}-k_{\nu _i} \right)}-1}&		\kappa _{\nu _i}\ne 0,\iota _{\tau _i}=0\\	\frac{1}{N}\sum_{\bar{n}=0}^{N-1}{e^{\small{j2\pi d_{\bar{n},n}\iota _{\tau _i}}}}&		\kappa _{\nu _i}\ne 0,\iota _{\tau _i}\ne 0\\	\,\,       \hspace{25pt}\times e^{\small{\frac{-j2\pi}{N}}\left( m-n+2Nc_1l_{\tau _i}-k_{\nu _i} \right) \bar{n}}&		\\\end{cases}.
		\end{aligned}
	\end{equation}
%	Note that the phase of the delay and Doppler coupling that is ignored is $e^{\small{\frac{-j2\pi}{N}k_{\nu _i}l_{\tau _i}}}$ in $\alpha \left( m,n,k_{\nu _i},l_{\tau _i} \right) $.
\end{theorem} 
\begin{IEEEproof}[Proof]
	The proof is given in Appendix A.
\end{IEEEproof}
\newtheorem{corollary}{\bf Corollary}
\renewcommand{\thecorollary}{\textup{\arabic{corollary}}}
%\begin{corollary}
%	Ignoring the DD coupling phase $e^{\small{\frac{-j2\pi}{N}k_{\nu _i}l_{\tau _i}}}$ in \eqref{alpha}, 
%	then \eqref{y_m_2} degenerates into the classical AFDM \cite{Bemani_3}. In other words, our work in \textbf{Theorem} \ref{thm1} can be considered as a enhanced AFDM when the DD coupling phase is considered. 
%\end{corollary}

$\mathcal{G} (m,n,k_{\nu _i},l_{\tau _i})$ is the AFDM basis function used for subsequent CE. Rewriting \eqref{y_m_2} into another matrix form, we have
\begin{equation}
	\boldsymbol{y}=\boldsymbol{H}_{\mathrm{eff}}\boldsymbol{x}+\boldsymbol{w},
\end{equation}
where the effective channel matrix $\boldsymbol{H}_{\mathrm{eff}}$ is the matrix form of $H_\mathrm{eff}\left( m,n \right)$ in the DAFT domain. 
In addition, the permutation matrix can typically model the integer delay in the time-domain channel matrix $\boldsymbol{H}_{\mathrm{T}}$.
However, this method fails for fractional delay. Since $\boldsymbol{H}_{\mathrm{eff}}=\boldsymbol{AH}_{\mathrm{T}}\boldsymbol{A}^{\mathrm{H}}$, $\boldsymbol{H}_{\mathrm{T}}$ is given by $\boldsymbol{H}_{\mathrm{T}}=\boldsymbol{A}^{\mathrm{H}}\boldsymbol{H}_{\mathrm{eff}}\boldsymbol{A}$.

\subsection{Analysis of ICCI and the Channel in the DAFT Domain}
Since $\alpha \left( m,n,k_{\nu _i},l_{\tau _i} \right)$ is a phase factor, the characteristics of $\boldsymbol{H}_{\mathrm{eff}}$ are determined by $\mathcal{F} \left( m,n,l_{\tau _i},k_{\nu _i} \right)$.

\textit{1) Integer-delay-integer-Doppler (IDID) Channels:} In this case, $l_{\tau _i}=\bar{l}_{\tau _i}$, $k_{\nu _i}=\bar{k}_{\nu _i}$.
%$\mathcal{F} \left( m,n,k_{\nu _i},l_{\tau _i} \right) $ will be simplified to
%\begin{equation}
%	\mathcal{F} \left( m,n,k_{\nu _i},l_{\tau _i} \right) =\left\{ \begin{matrix}	1,&		\left[ m-n+2Nc_1l_{\tau _i}-k_{\nu _i} \right] _N=0\\	0,&		\mathrm{otherwise}\\\end{matrix} \right. ,
%\end{equation}
%where $\left[ \cdot \right] _N$ is the mod-$N$ operation.
Since $\mathcal{F} \left( m,n,l_{\tau _i},k_{\nu _i} \right)$ simplifies to $\frac{1}{N}\delta \left( \left[ m-n+2Nc_1l_{\tau _i}-k_{\nu _i} \right] _N \right)$, \eqref{y_m_2} will be simplified to
\begin{equation}\label{IDD}
	y\left[m\right]=\sum_{i=1}^P{h_i\alpha (m,n,k_{\nu _i},l_{\tau _i})x\left[n\right]}+w\left[n\right],
\end{equation}
where $n=\left[ m+2Nc_1l_{\tau _i}-k_{\nu _i} \right] _N$. 
For the $i$-th path, since only the transmitted signal $x\left[ \left[ m+2Nc_1l_{\tau _i}-k_{\nu _i} \right] _N \right] $ constitutes $y\left[ m \right] $, there is no ICCI for the IDID channels.

\textit{2) IDFD Channels:} In this case, $l_{\tau _i}=\bar{l}_{\tau _i}$, $k_{\nu _i}=\bar{k}_{\nu _i}+\kappa _{\nu _i}$. 
Since $\iota _{\tau _i}= 0$, we have
\begin{equation}\label{F_IDFD}
	\begin{aligned}
		\mathcal{F} \left( m,n,l_{\tau _i},k_{\nu _i} \right) &=\frac{e^{\small{-j2\pi}\beta}-1}{Ne^{\small{\frac{-j2\pi}{N}}\beta}-1}\\
		&=\frac{1}{N}e^{-j(N-1)\pi \frac{\beta}{N}}\frac{\sin \left( \pi \beta \right)}{\sin \left( \frac{\pi \beta}{N} \right)},
	\end{aligned}
\end{equation}
where $\beta =m-n+2Nc_1l_{\tau _i} -k_{\nu _i}$.
Due to the existence of fractional Doppler $\kappa _{\boldsymbol{\nu }_{\boldsymbol{i}}}$, $\mathcal{F} (m,n,k_{\nu _i},l_{\tau _i})$ will not be zero for a given $m$ and any $n$. Some works \cite{P. Raviteja_1,Bemani_3} have shown that $\left| \mathcal{F} (m,n,k_{\nu _i},l_{\tau _i}) \right|$ has a peak at $n=\left[ m+2Nc_1l_{\tau _i}-\bar{k}_{\nu _i} \right] _N$ and decreases significantly as $n$ moves away from $\left[ m+2Nc_1l_{\tau _i}-\bar{k}_{\nu _i} \right] _N$.
Therefore, we only need to consider the $2\xi +1$ principal values of $\mathcal{F} (m,n,k_{\nu _i},l_{\tau _i})$ around the peak of $\left[ m+2Nc_1l_{\tau _i}-\bar{k}_{\nu _i} \right] _N$, i.e., $\left[ m+2Nc_1l_{\tau _i}-\bar{k}_{\nu _i}-\xi  \right] _N\leqslant n\leqslant \left[ m+2Nc_1l_{\tau _i}-\bar{k}_{\nu _i}+\xi  \right] _N$. Based on this good approximation, we can re-express $y\left[ m \right] $ in \eqref{y_m_2} as
\begin{equation}\label{IDFD_y_m}
	\begin{aligned}
		y\left[ m \right] \approx& \sum_{i=1}^P{\sum_{n=\left[ m+2Nc_1l_{\tau _i}-\bar{k}_{\nu _i}-\xi \right] _N}^{\left[ m+2Nc_1l_{\tau _i}-\bar{k}_{\nu _i}+\xi \right] _N}{x \left[n\right] h_i\mathcal{G} (m,n,k_{\nu _i},l_{\tau _i})}}\\\approx& \sum_{i=1}^P{\sum_{q=-\xi}^{\xi}{x \left[n_q\right] h_i\alpha (m,n_q,k_{\nu _i},l_{\tau _i})}}\\&\times\frac{e^{-j2\pi (q-\kappa _{\nu _i})}-1}{Ne^{-j\frac{2\pi}{N}(q-\kappa _{\nu _i})}-N},
	\end{aligned}
\end{equation}
where $n_q=\left[ m-q+2Nc_1l_{\tau _i}-\bar{k}_{{\nu }_{{i}}} \right] _N$. 
From \eqref{IDFD_y_m}, it is evident that $y\left[ m \right] $ is approximately a linear combination of $P\left( 2\xi +1 \right) $ transmitted signals. For the $2\xi +1$ transmitted signals in the $i$-th path, only $x\left[ n_{q\left| q=0 \right.} \right] $ is the main source of $y\left[ m \right] $, and the other $2\xi $ signals can be regarded as interference. This interference is caused by the chirp-subcarriers near the $n_{q\left| q=0 \right.}$-th chirp-subcarrier, so it is called ICCI.

\textit{3) FDFD Channels:} In this case, $l_{\tau _i}=\bar{l}_{\tau _i}+\iota _{\tau _i}$, $k_{\nu _i}=\bar{k}_{\nu _i}+\kappa _{\nu _i}$.
%Since both $2Nc_1\iota _{\tau _i}$ and $2Nc_1\iota _{\tau _i}-\lfloor 2Nc_1\iota _{\tau _i} \rceil -\kappa _{\nu _i}$ may also have the possibility of generating fractional components,
Since $\iota _{\tau _i}\ne 0$, we have
	\begin{equation}\label{F_FDFD}
		\mathcal{F} \left( m,n,k_{\nu _i},l_{\tau _i} \right) =\frac{1}{N}\sum_{\bar{n}=0}^{N-1}{e^{\small{\frac{-j2\pi}{N}}\eta _{m,n}^{\left( i \right)}\bar{n}}e^{j2\pi \left( d_{\bar{n},n}-2\bar{n}c_1 \right) \iota _{\tau _i}},}
	\end{equation}
	where $\eta _{m,n}^{\left( i \right)}=m-n-\chi _i-\kappa _{\nu _i}$, $\chi _i=-2Nc_1\bar{l}_{\tau _i}+\bar{k}_{\nu _i}$ is the equivalent shift of the $i$-th path. Since $\left(d_{\bar{n},n}-2\bar{n}c_1 \right) \iota _{\tau _i}$ is small, it does not affect the peak position. For the $i$-th path, the condition for the occurrence of the main peak of $\left| \mathcal{F} \left( m,n,k_{\nu _i},l_{\tau _i} \right) \right|$ is
	\begin{equation}\label{main_peak}
		m=m_\mathrm{m-peak}^{\left(i\right)}=\left[ n+\chi _i \right] _N.
	\end{equation}
	In addition, $\left| \mathcal{F} \left( m,n,k_{\nu _i},l_{\tau _i} \right) \right|$ displays some high-level local peaks. The positions of these local peaks are
	\begin{equation}\label{side_peak}
		m_{\mathrm{l}-\mathrm{peak}}^{\left( i \right)}=m_{\mathrm{m}-\mathrm{peak}}^{\left( i \right)}\pm 2gNc_1,
	\end{equation}
	where $g$ is a small positive integer. In addition, as $m_{\mathrm{l}-\mathrm{peak}}^{\left( i \right)}$ moves away from $m_{\mathrm{m}-\mathrm{peak}}^{\left( i \right)}$, its local peaks decrease in magnitude. Therefore, for FDFD channels, these local peaks will lead to more severe ICCI. Simultaneously, local peaks in all paths further exacerbate the inter-path interference.

Based on the above analysis, since $\left| \mathcal{F} \left( m,n,k_{\nu _i},l_{\tau _i} \right) \right|$ contains non-negligible local peaks, the I/O relationship cannot be approximated and can only be expressed as \eqref{y_m_2}. In addition, $N-1$ transmitted signals will generate ICCI for $y\left[m\right].$

	\subsection{Low-Complexity Channel Matrix Construction}
	The rapid and low-complexity construction of channel matrices enables efficient system simulations, reduces computational overhead, and facilitates real-time signal processing, which is particularly important for high-dimensional or large-scale systems.

The complexity of constructing $\boldsymbol{H}_{\mathrm{eff}}$ mainly comes from the construction of the $\mathcal{F} \left( m,n,l_{\tau _i},k_{\nu _i} \right)$.
	For both IDID and IDFD channels, $\mathcal{F} \left( m,n,l_{\tau _i},k_{\nu _i} \right)$ can be converted into the simple form of the summation-free version in \eqref{F_2}. However, for FDFD channels, due to the presence of $d_{\bar{n},n}\iota _{\tau _i}$, $\mathcal{F} \left( m,n,l_{\tau _i},k_{\nu _i} \right)$ cannot be further simplified. Using the element-wise summation (ES) method to construct the channel matrix leads to high complexity, especially for large $N$.

\textit{1) The FFT-based Method:} 

We rewrite  $\mathcal{F} \left( m,n,l_{\tau _i},k_{\nu _i} \right)$ as
	\begin{equation}
		\begin{aligned}
			\mathcal{F} \left( m,n,l_{\tau _i},k_{\nu _i} \right) &=\sum_{\bar{n}=0}^{N-1}{e^{\small{\frac{-j2\pi}{N}}\left( \left( 2Nc_1l_{\tau _i}-k_{\nu _i}-n \right) \bar{n}-Nd_{\bar{n},n}\iota _{\tau _i} \right)}}\\ &\hspace{30pt}\times e^{\small{\frac{-j2\pi}{N}}m\bar{n}}\\&=\sum_{\bar{n}=0}^{N-1}{u\left( \bar{n},n \right) e^{\small{\frac{-j2\pi}{N}}m\bar{n}}},
		\end{aligned}
	\end{equation}
	where $u\left( \bar{n},n \right) =e^{\small{\frac{-j2\pi}{N}}\left( \left( 2Nc_1l_{\tau _i}-k_{\nu _i}-n \right) \bar{n}-Nd_{\bar{n},n}\iota _{\tau _i} \right)}$. This is exactly the DFT Transform. Utilizing the FFT enables the rapid construction of $\mathcal{F} \left( m,n,l_{\tau _i},k_{\nu _i} \right)$. 

\textit{2) The Segment-wise Summation (SS) Method:}

Considering the special structure of $d_{\bar{n},n}$, $\left| \mathcal{F} (m,n,k_{\nu _i},l_{\tau _i}) \right|$ can be simplified into a relatively simple segment-wise summation form. $d_{\bar{n},n}$ exhibits a staircase structure. For a given fixed $n$, the staircase has $R_n$ levels. In addition, let the starting index, width, and value of each level are $\bar{n}_{r,n}$,  $L_{r,n}$, and $d_{r,n}$, respectively. Therefore, \eqref{F_FDFD} can be rewritten as
	\begin{equation}
		\begin{aligned}
			\mathcal{F} \left( m,n,l_{\tau _i},k_{\nu _i} \right) &=\frac{1}{N}\sum_{r=0}^{R_n-1}{e^{j2\pi d_{r,n}\iota _{\tau _i}}\sum_{\bar{n}=\bar{n}_{r,n}}^{\bar{n}_{r,n}+L_{r,n}-1}{e^{\small{\frac{-j2\pi}{N}}\beta \bar{n}}}}
			\\&=\frac{1}{N\sin \left( \frac{\pi \beta}{N} \right)}\sum_{r=0}^{R_n-1}{\sin \left( \frac{\pi}{N}\beta L_{r,n} \right) e^{j\varphi _{r,n}}},
		\end{aligned}
	\end{equation}
	where $\varphi _{r,n}=2\pi d_{r,n}\iota _{\tau _i}-\small{\frac{2\pi \beta}{N}\bar{n}_{r,n}^{\mathrm{c}}}$, $\bar{n}_{r,n}^{\mathrm{c}}=\bar{n}_{r,n}+\frac{L_{r,n}-1}{2}$ is the central index of the $r$-th segment, $\beta =m-n+2Nc_1l_{\tau _i} -k_{\nu _i}$. In addition, the mean $R$ of $R_n$ is generally $R\approx 2Nc_1$.

	The ES method directly calculates each matrix element through explicit summation over all $\bar{n}$, resulting in a cubic complexity of $\mathcal{O} \left( N^3 \right) $. The FFT-based method exploits the inherent DFT structure of the summation over $\bar{n}$ in $\mathcal{F} \left( m,n,l_{\tau _i},k_{\nu _i} \right)$. By applying FFT to accelerate the computation, the overall complexity is reduced to $\mathcal{O} \left( N^2\log N \right) $. By leveraging the staircase structure of $d_{\bar{n},n}$, the SS method decomposes the summation over $n$ into a series of closed expressions from each segment.
	Thus, the SS method effectively reduces the summation size from $N$ to $R$, yielding complexity $\mathcal{O} \left( N^2 R \right) $, far below the complexity of ES method.

\subsection{SINR}
For the receiver, linear minimum mean square error (LMMSE) equalization is used for $\boldsymbol{y}$ to mitigate the effects of noise and interference, i.e., $\hat{\boldsymbol{x}}=\boldsymbol{Wy}$, where $\hat{\boldsymbol{x}}$ is the estimate of $\boldsymbol{x}$, $\boldsymbol{W}=\left( \boldsymbol{H}_{\mathrm{eff}}^{\mathrm{H}}\boldsymbol{H}_{\mathrm{eff}}+\sigma ^2\boldsymbol{I}_N \right) ^{-1}\boldsymbol{H}_{\mathrm{eff}}^{\mathrm{H}}$. The signal of the $n$-th chirp-subcarrier is
\begin{equation}
	\hat{x}\left[ n \right] =T\left[ n,n \right] x\left[ n \right] +\sum_{m=0,m\ne n}^N{T\left[ n,m \right] x\left[ m \right]}+\tilde{w}\left[ n \right] ,
\end{equation}
where $n\in \left\{ 0,1,\cdots ,N-1 \right\} $, $\boldsymbol{T}=\boldsymbol{WH}_{\mathrm{eff}}$. Therefore, the SINR on the $n$-th chirp-subcarrier is expressed as
\begin{equation}
	\mathrm{SINR}_n=\small{\frac{\left| T\left[ n,n \right] \right|^2}{\sum_{m=0,m\ne n}^{N-1}{\left| T\left[ n,m \right] \right|^2}+\sigma ^2\left\| \left\{ \boldsymbol{W} \right\} _{n,:} \right\| ^2}},
\end{equation}
where $\left\{ \cdot \right\} _{n,:}$ is the $n$-th row of the matrix.

\subsection{Effective SINR Loss}
For the IDID channels, according to the I/O relationship of \eqref{IDD}, the corresponding peak power of the $i$-th path is
\begin{equation}\label{hi_p_IDD}
	P_{\mathrm{IDD}}=\left| h_i \right|^2.
\end{equation}
It can be seen that there is no SINR loss in this case.
%However, the fractional channels will inevitably lead to SINR loss.

For IDFD channels, based on subsection \Rmnum{3}-B-2), we have
%$\left| \mathcal{G}\left( m,n,k_{\nu _i},l_{\tau _i} \right) \right|$ can be expressed as
%\begin{equation}\label{h_w_abs}
%	\left| \mathcal{G}\left( m,n,k_{\nu _i},l_{\tau _i} \right) \right|=\left| \frac{\sin \left( \pi \left( \psi -\kappa _{\nu _i} \right) \right)}{N\sin \left( \frac{\pi (\psi -\kappa _{\nu _i})}{N} \right)} \right|,
%\end{equation}
%where $\psi =\left[ m-n+2Nc_1l_{\tau _i}-\bar{k}_{\nu _i} \right] _N$ is integer. As mentioned before, $\left| h_w\left( m,n,k_{\nu _i},l_{\tau _i} \right) \right|$ has a peak at $\psi =0$, i.e.,
\begin{equation}
	\left| \mathcal{G}\left( m,n,k_{\nu _i},l_{\tau _i} \right) \right|\leqslant \left| \frac{\sin \left( \pi \kappa _{\nu _i} \right)}{N\sin \left( \frac{\pi \kappa _{\nu _i}}{N} \right)} \right|.
\end{equation}
Therefore, the corresponding peak power of the $i$-th path is
\begin{equation}
	P_{\mathrm{IDFD}}=\left| h_i \right|^2\left| \frac{\sin \left( \pi \kappa _{\nu _i} \right)}{N\sin \left( \frac{\pi \kappa _{\nu _i}}{N} \right)} \right|^2.
\end{equation}
Compared with \eqref{hi_p_IDD}, the SINR loss in the IDFD channels is
\begin{equation}\label{L_SNR_IDFD}
	\begin{aligned}
		L_{\mathrm{SINR}}^\mathrm{IDFD}=&20\log _{10}\left| \frac{\sin \left( \pi \kappa _{\nu _i} \right)}{N\sin \left( \frac{\pi \kappa _{\nu _i}}{N} \right)} \right|^2\approx 20\log _{10}\left| \frac{\sin \left( \pi \kappa _{\nu _i} \right)}{\pi \kappa _{\nu _i}} \right|^2,
	\end{aligned}
\end{equation}
where ${\pi \kappa _{\nu _i}}/{N}$ approaches zero, resulting in $\sin \left( {\pi \kappa _{\nu _i}}/{N} \right) \approx {\pi \kappa _{\nu _i}}/{N}$. In general, $N\gg \pi k_{\nu _i}$ is satisfied, therefore this approximation holds.  Moreover, according to \eqref{L_SNR_IDFD}, increasing $N$ cannot effectively reduce the SINR loss.

For FDFD channels, based on \eqref{F_FDFD}-\eqref{main_peak}, the peak value $\mathcal{F} _{\max}$ of $\left| \mathcal{F} \left( m,n,k_{\nu _i},l_{\tau _i} \right) \right|$ can be approximated as
		\begin{equation}\label{F_max_approx}
		\mathcal{F} _{\max}\approx \left| \frac{1}{N}\sum_{\bar{n}=0}^{N-1}{e^{\small{\frac{-j2\pi}{N}}\eta _{m_{\mathrm{m}-\mathrm{peak}}^{\left( i \right)},n}^{\left( i \right)}\bar{n}}} \right|\approx \left| \frac{\sin \left( \pi \kappa _{\nu _i} \right)}{\pi \kappa _{\nu _i}} \right|.
	\end{equation}
%	we have
%	\begin{equation}
%		\left| \mathcal{G} \left( m,n,l_{\tau _i},k_{\nu _i} \right) \right|\leqslant \mathcal{F} _{\max},
%	\end{equation}
%	\begin{equation}\label{F_max}
%		\begin{aligned}
%		\mathcal{F} _{\max}&=\left| \frac{1}{N\sin \left( \frac{\pi \beta _{\mathrm{m}-\mathrm{peak}}}{N} \right)}\sum_{r=0}^{R-1}{\sin \left( \frac{\pi}{N}\beta _{\mathrm{m}-\mathrm{peak}}L_r \right) e^{j\varphi _r}} \right|\\&\approx \left| \frac{1}{\pi R\iota _{\tau _i}}\sum_{r=0}^{R-1}{\sin \left( \pi\iota _{\tau _i} \right) e^{j\tilde{\varphi}_r}} \right|,
%		\end{aligned}
%	\end{equation}
%	where $\tilde{\varphi}_r=2\pi d_r\iota _{\tau _i}-\small{\frac{2\pi \iota _{\tau _i}}{L}\bar{n}_{r}^{\mathrm{c}}}$. The approximation in \eqref{F_max} is similar to that in \eqref{L_SNR_IDFD}.
	 Therefore, increasing $N$ does not reduce the SINR loss in the FDFD channel.
	 The SINR loss in the FDFD channels is
	 \begin{equation}\label{L_SNR_FDFD}
	 	\begin{aligned}
	 		L_{\mathrm{SINR}}^{\mathrm{FDFD}}&=20\log _{10}\left| \mathcal{F} \left( m_{\mathrm{m}-\mathrm{peak}}^{\left( i \right)},n,k_{\nu _i},l_{\tau _i} \right) \right|^2.
	 	\end{aligned}
	 \end{equation}

%similar to the previous derivation, 
%%the peak value of $\left| \mathcal{G}\left( m,n,k_{\nu _i},l_{\tau _i} \right) \right|$ can be expressed as
%%\begin{equation}
%%	\\\left| \mathcal{G}\left( m,n,k_{\nu _i},l_{\tau _i} \right) \right|\leqslant \left| \frac{\sin \left( \pi \gamma_i \right)}{N\sin \left( \frac{\pi \gamma_i}{N} \right)} \right|.
%%\end{equation}
%the peak power corresponding to the $i$-th path is
%\begin{equation}
%	P_{\mathrm{FDFD}}=\left| h_i \right|^2\left| \frac{\sin \left( \pi \gamma_i \right)}{N\sin \left( \frac{\pi \gamma_i}{N} \right)} \right|^2.
%\end{equation}
%Compared with \eqref{hi_p_IDD}, the SINR loss is
%\begin{equation}\label{L_SNR_FDD}
%	\begin{aligned}
%		L_{\mathrm{SINR}}^\mathrm{FDFD}=&20\log _{10}\left| \frac{\sin \left( \pi \gamma_i \right)}{N\sin \left( \frac{\pi \gamma_i}{N} \right)} \right|^2\approx 20\log _{10}\left| \frac{\sin \left( \pi \gamma_i \right)}{\pi \gamma_i} \right|^2.
%	\end{aligned}
%\end{equation}

%\begin{figure}[h]
%	\centerline{\includegraphics[width = 7cm ]{./fig/L_SNR/L_SNR.eps}}
%	\caption{Effective SNR loss vs. $\iota _{\tau _i}$ and $\kappa _{\nu _i}$.}
%	\label{L_SNR}
%\end{figure}
\begin{figure}[t]
	\centerline{\includegraphics[width=\columnwidth ]{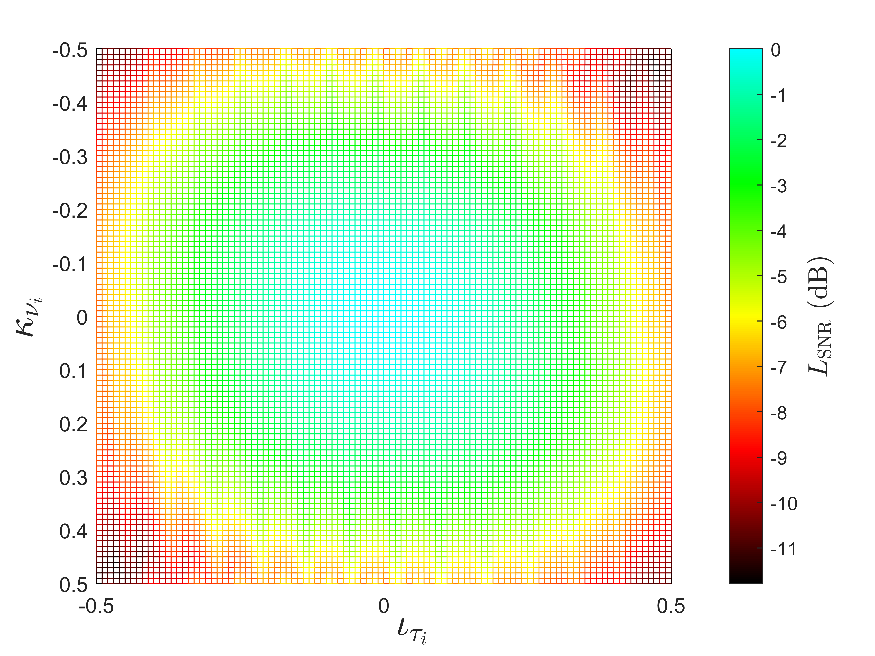}}
	\caption{Planform of SINR loss.}
	\label{L_SNR_fs}
\end{figure}
%\begin{figure}[h]
%	\centerline{\includegraphics[width = 7cm ]{./fig/L_SNR/L_SNR_N.eps}}
%	\caption{SINR loss vs. $N$.}
%	\label{L_SNR_N}
%\end{figure}
Based on \eqref{L_SNR_FDFD}, Fig. \ref{L_SNR_fs} shows the effective SINR loss for different fractional delays and fractional Dopplers at $N=64$. Increasing $\iota_{\tau_i}$ or $\kappa_{\nu _i}$ results in more severe SINR loss. The maximum SINR loss is about 11.788 dB. Moreover, according to \eqref{L_SNR_IDFD} and \eqref{F_max_approx}, increasing $N$ cannot effectively reduce the SINR loss.
%In addition, since $2Nc_1\iota _{\tau _i}$ can make the fractional part of the delay small enough, we can see that the $\iota _{\tau _i}$ has almost no effect on SINR loss when $\kappa _{\nu _i}=0$.

%\theoremstyle{remark}
%\newtheorem{remark2}{\bf Remark}
%\begin{remark}
%%	The fractional delay may reduce rather than increase SINR loss, which can also be seen from Fig. \ref{L_SNR_fs}. The reason for this is that when the signs of the fractional delay and fractional Doppler are the same, the fractional parts can be offset to reduce the SINR loss.
%	As shown in Fig. \ref{L_SNR_fs}, the joint effects of fractional delay and fractional Doppler may result in a reduction of $\gamma_i$ to alleviate the SINR loss.
%\end{remark}

%Fig. \ref{L_SNR0} is a profile of SNR loss for $\iota _{\tau _i}=0$ or $\kappa _{\nu _i}=0$ at $N=64$.
%Fig. \ref{L_SNR_N} shows the effect of fractional channels on SINR loss for different $N$. It can also be seen that increasing $N$ cannot effectively reduce the SINR loss for a given fractional parameter. 
%Increasing $N$ can get a larger $l _{\tau _i}$, and $k _{\nu _i}$ is not affected, but $\iota _{\tau _i}$ and $\kappa _{\nu _i}$ still exist.
%Even with a very large $B$, the loss may reach about 7.84 dB. 

\section{MF and MF-GFS CE Schemes}
%Most existing works assume that the system bandwidth $B$ is large enough so that the delay resolution $\frac{1}{N\varDelta f}$ can approximate the path delay to the nearest sampling point. However, in practice, the limited time-frequency resources, i.e., the bandwidth and duration corresponding to an AFDM frame, will inevitably lead to the delay and Doppler resolution being limited by the Rayleigh resolution. To achieve good enough performance, we need to consider the impact of fractional delay.

In this section, we propose two low-complexity methods for estimating path parameters, by sequentially estimating the parameters of each individual path.
Note that the chirp characteristic of separating different paths by $c_1$ is leveraged in the proposed CE schemes. Due to the fact that the chirp characteristic is invalidated by $c_1=0$ in OFDM systems, the method proposed cannot be directly applied to OFDM systems.

\subsection{Problem Description}
According to \eqref{y_m_2}-\eqref{F_2}, the system model is
\begin{equation}
	\boldsymbol{y}={\boldsymbol{A}}\left( \boldsymbol{l}_{\tau},\boldsymbol{k}_{\nu} \right) \boldsymbol{h}+\boldsymbol{w},
\end{equation}
where $\boldsymbol{y}\in \mathbb{C} ^{N\times 1}$ and $\boldsymbol{w}\sim \mathcal{C} \mathcal{N} \left( 0,\sigma ^2\boldsymbol{I}_N \right) \in \mathbb{C} ^{N\times 1}$ are vector forms of $y\left( m \right) $ and $w\left( m \right) $ respectively, vectors $\boldsymbol{l}_{\tau}=\left[ l_{\tau _1},\cdots ,l_{\tau _P} \right] ^{\mathrm{T}}\in \mathbb{R} ^{P\times 1}$, $\boldsymbol{k}_{\nu}=\left[ k_{\nu _1},\cdots ,k_{\nu _P} \right] ^{\mathrm{T}}\in \mathbb{R} ^{P\times 1}$ and $\boldsymbol{h}=\left[ h_1,\cdots ,h_P \right] ^{\mathrm{T}}\in \mathbb{C} ^{P\times 1}$ contain the normalized delay, normalized Doppler and channel gain, respectively. ${\boldsymbol{A}}\left( \boldsymbol{l}_{\tau},\boldsymbol{k}_{\nu} \right) \in \mathbb{C} ^{N\times P}$ can be expressed as
\begin{equation}
	{\boldsymbol{A}}\left( \boldsymbol{l}_{\tau},\boldsymbol{k}_{\nu} \right) =\left[ {\boldsymbol{a}}\left( l_{\tau _1},k_{\nu _1} \right) ,\cdots ,{\boldsymbol{a}}\left( l_{\tau _P},k_{\nu _P} \right) \right] ,
\end{equation}
where ${\boldsymbol{a}}\left( l_{\tau _i},k_{\nu _i} \right) \in \mathbb{C} ^{N\times 1}$, the $m$-th entry of ${\boldsymbol{a}}\left( l_{\tau _i},k_{\nu _i} \right) $ is
\begin{equation}
	\left\{ {\boldsymbol{a}}\left( l_{\tau _i},k_{\nu _i} \right) \right\} _m=\sum_{n=0}^{N-1}{x\left( n \right) \mathcal{G} \left( m,n,k_{\nu _i},l_{\tau _i} \right)},
\end{equation}
where $m\in \left\{ 0,\cdots ,N-1 \right\} $.
%The purpose of PE is to estimate normalized Doppler $\boldsymbol{k}_{\nu}$, normalized delay $\boldsymbol{l}_{\tau}$ and channel coefficient $\boldsymbol{h}$ based on $\tilde{\boldsymbol{A}}\left( \boldsymbol{l}_{\tau},\boldsymbol{k}_{\nu} \right) $ and received signal $\boldsymbol{y}$.
It can be seen that ${\boldsymbol{A}}\left( \boldsymbol{l}_{\tau},\boldsymbol{k}_{\nu} \right) $ contains data, and channel information including $k_{\nu _i}$, $l_{\tau _i}$,$P$, which need to be estimated.

\begin{figure}[t]
	\centerline{\includegraphics[width=\columnwidth ]{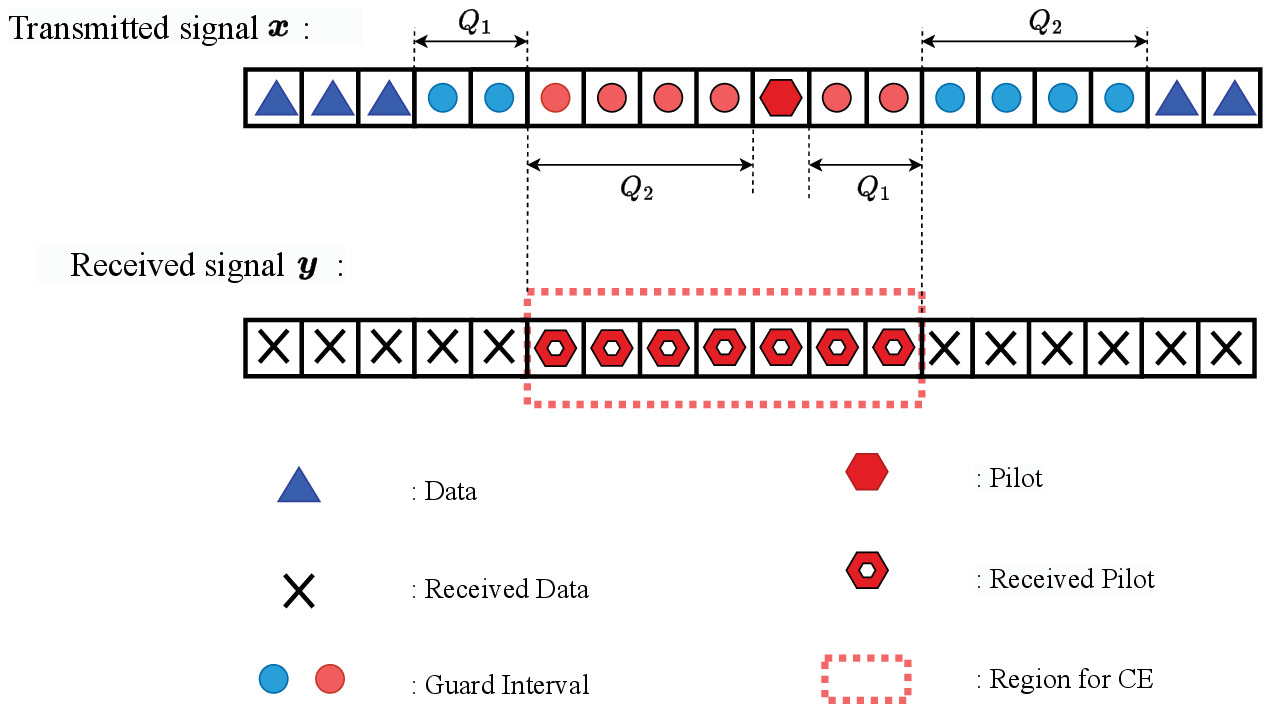}}
	\caption{Single pilot pattern for AFDM systems.}
	\label{Pilot_Pattern}
\end{figure}

\textit{1) Pilot Pattern:}
Pilots may be inserted into the DAFT domain for efficient CE.
%The unknown data symbols problem can be addressed through pilot symbols.
Due to the sparsity and compactness of the channel in the DAFT domain, a single pilot can be considered.
The arrangement of pilot and data is shown in Fig. \ref{Pilot_Pattern}, which can be expressed as
\begin{equation}
	\begin{aligned}
		x\left[ n \right] =\begin{cases}	x_p,&		n=n_p\\	0,&		n_p-Q\leqslant n\leqslant n_p+Q,n\ne n_p\\	x_{\mathrm{data}},&		\mathrm{otherwise}\\\end{cases},
	\end{aligned}
\end{equation}
where $x_p$ and $x_{\mathrm{data}}$ are single pilot and data, respectively, $n_p$ is the pilot index in the DAFT domain grid, $Q\triangleq (l_{\max}+1)(2(k_{\max}+\xi )+1)-1$ due to the structure of $\mathcal{G}\left( m,n,k_{\nu _i},l_{\tau _i} \right) $ and the analysis in Section \Rmnum{3}. The DAFT domain grid within the range $n_p-Q\leqslant n\leqslant n_p+Q, n\ne n_p$ remains null as a guard interval to reduce the mutual interference between data and pilot. 
Since $\bar{l}_{\tau _i}$ in the equivalent shift $\chi _i=-2Nc_1 \bar{l}_{\tau _i} +\bar{k}_{\nu _i}$ is non-negative, the guard intervals on the two sides of the pilot have different lengths. Thus, the right and left guard intervals are of lengths $Q_1=k_{\max}+\xi $ and $Q_2=Q-Q_1 $, respectively. In addition, the two colored guard intervals in Fig. \ref{Pilot_Pattern} come from pilot and data, respectively.

%Usually, only the pilot symbol part in the matrix $\tilde{\boldsymbol{A}}\left( \boldsymbol{l}_{\tau},\boldsymbol{k}_{\nu} \right) $ is used for PE. And the remaining data symbol part of $\tilde{\boldsymbol{A}}\left( \boldsymbol{l}_{\tau},\boldsymbol{k}_{\nu} \right) $ is regarded as the model error, which is absorbed into the noise.
%In addition, benefiting from the compactness of channel in the DAFT domain, 
%%specifically $\tau _i\in \left[ 0,\tau _{\max} \right] $ and $\nu _i\in \left[ -\nu _{\max},\nu _{\max} \right] $, 
%the main pilot response at the receiver will be confined to a small region. 
%Therefore, as shown in the red box region in Fig. \ref{Pilot_Pattern}, the received signal $\boldsymbol{y}$ within the range $n_p-Q_2\leqslant m\leqslant n_p+Q_1$ is used for PE to reduce the computational load.

\textit{2) CE Model:} Since the main pilot response in $\boldsymbol{y}$ is distributed in the region for CE, truncating the $\boldsymbol{y}$ is intended to reduce complexity. In particular, we employ the received signals $y\left[m\right]$ in the range of $n_p-Q_2\leqslant m\leqslant n_p+Q_1$ to estimate the channel. Therefore, $y\left[ m \right]$ in \eqref{y_m_2} is simplified to
\begin{equation}\label{y_m_p}
	y\left[ m \right] =\sum_{i=1}^P{h_i\mathcal{G} \left( m,n_p,k_{\nu _i},l_{\tau _i} \right) x_p}+z\left[ n \right],
\end{equation}
where $z\left[ n \right]$ is the new noise.
%which is the sum of the noise $w\left( n \right)$ and the data symbol part in $\tilde{\boldsymbol{A}}\left( \boldsymbol{l}_{\tau},\boldsymbol{k}_{\nu} \right) $.
The matrix form corresponding to \eqref{y_m_p} is
\begin{equation}\label{y_T}
	\boldsymbol{y}_{\mathrm{T}}=x_p\boldsymbol{A}_{\mathrm{T}}\left( \boldsymbol{l}_{\tau},\boldsymbol{k}_{\nu} \right) \boldsymbol{h}+\boldsymbol{z}_{\mathrm{T}},
\end{equation}
where the subscript $\mathrm{T}$ indicates truncation, i.e., $n_p-Q_2\leqslant m\leqslant n_p+Q_1$. In the truncated $\boldsymbol{A}_{\mathrm{T}}\left( \boldsymbol{l}_{\tau},\boldsymbol{k}_{\nu} \right) \in \mathbb{C} ^{\left( Q+1 \right) \times P}$, $\boldsymbol{a}_{\mathrm{T}}\left( l_{\tau _i},k_{\nu _i} \right) \in \mathbb{C} ^{\left( Q+1 \right) \times 1}$ is given by
\begin{equation}
	\begin{aligned}
		\left\{ \boldsymbol{a}_{\mathrm{T}}\left( l_{\tau _i},k_{\nu _i} \right) \right\} _m=&\mathcal{G}\left( m,n_p,k_{\nu _i},l_{\tau _i} \right).
	\end{aligned}
\end{equation}
\textit{3) Decoupling Multipath Estimation:}
Based on \eqref{y_T}, the joint maximum likelihood estimation (JMLE) of $\left( \boldsymbol{l}_{\tau},\boldsymbol{k}_{\nu},\boldsymbol{h} \right) $ can be expressed as
\begin{equation}\label{JMLE}
	\left( \hat{\boldsymbol{l}}_{\tau},\hat{\boldsymbol{k}}_{\nu},\hat{\boldsymbol{h}} \right) =\mathrm{arg}\underset{\left( \boldsymbol{l}_{\tau},\boldsymbol{k}_{\nu},\boldsymbol{h} \right)}{\min}\left\| \boldsymbol{y}_{\mathrm{T}}-x_p\boldsymbol{A}_{\mathrm{T}}\left( \boldsymbol{l}_{\tau},\boldsymbol{k}_{\nu} \right) \boldsymbol{h} \right\| ^2.
\end{equation}
According to \cite{Khan_1}, $\left( \hat{\boldsymbol{l}}_{\tau},\hat{\boldsymbol{k}}_{\nu} \right)$ can be expressed as \eqref{JMLE_lk}, as shown at the bottom of next page,
\begin{figure*}[b] % hb底部，ht为头部
	\centering % 公式居中
	\hrulefill % 添加一条水平线
	%	\vspace*{8pt} % 调整线与公式之间的距离
	\begin{equation}\label{JMLE_lk}
		\begin{aligned}
			\left( \hat{\boldsymbol{l}}_{\tau},\hat{\boldsymbol{k}}_{\nu} \right) =&\mathrm{arg}\underset{\left( \boldsymbol{l}_{\tau},\boldsymbol{k}_{\nu} \right)}{\max}\boldsymbol{y}_{\mathrm{T}}^{\mathrm{H}}\boldsymbol{A}_{\mathrm{T}}\left( \boldsymbol{l}_{\tau},\boldsymbol{k}_{\nu} \right) \left( \boldsymbol{A}_{\mathrm{T}}^{\mathrm{H}}\left( \boldsymbol{l}_{\tau},\boldsymbol{k}_{\nu} \right) \boldsymbol{A}_{\mathrm{T}}\left( \boldsymbol{l}_{\tau},\boldsymbol{k}_{\nu} \right) \right) ^{-1}\boldsymbol{A}_{\mathrm{T}}^{\mathrm{H}}\left( \boldsymbol{l}_{\tau},\boldsymbol{k}_{\nu} \right) \boldsymbol{y}_{\mathrm{T}}.
		\end{aligned}
	\end{equation}
\end{figure*}
$\hat{\boldsymbol{h}}$ is given by
\begin{equation}\label{h_ML_2}
	\begin{aligned}
		\hat{\boldsymbol{h}}=&\small{\frac{1}{x_p}}\left( \boldsymbol{A}_{\mathrm{T}}^{\mathrm{H}}\left( \hat{\boldsymbol{l}}_{\tau},\hat{\boldsymbol{k}}_{\nu} \right) \boldsymbol{A}_{\mathrm{T}}\left( \hat{\boldsymbol{l}}_{\tau},\hat{\boldsymbol{k}}_{\nu} \right) \right) ^{-1}\boldsymbol{A}_{\mathrm{T}}^{\mathrm{H}}\left( \hat{\boldsymbol{l}}_{\tau},\hat{\boldsymbol{k}}_{\nu} \right) \boldsymbol{y}_{\mathrm{T}}.
	\end{aligned}
\end{equation}

\newtheorem{theorem2}{\bf Theorem}

\begin{theorem}\label{thm2}
	For any $i\ne j$, $\left\| \boldsymbol{a}_{\mathrm{T}}\left( l_{\tau _i},k_{\nu _i} \right) \right\| ^2=1$,  
		\begin{equation}
			\left| \boldsymbol{a}_{\mathrm{T}}^{\mathrm{H}}\left( l_{\tau _i},k_{\nu _i} \right) \boldsymbol{a}_{\mathrm{T}}\left( l_{\tau _j},k_{\nu _j} \right) \right|\leqslant \epsilon,
		\end{equation}
		where $\epsilon =\underset{i\ne j}{\max}\left| \boldsymbol{a}_{\mathrm{T}}^{\mathrm{H}}\left( l_{\tau _i},k_{\nu _i} \right) \boldsymbol{a}_{\mathrm{T}}\left( l_{\tau _j},k_{\nu _j} \right) \right|$.
		\theoremstyle{remark}
		\newtheorem{remark1}{\bf Remark}
		\begin{remark}
		Since $\epsilon \ll 1$, $\boldsymbol{A}_{\mathrm{T}}^{\mathrm{H}}\left( \boldsymbol{l}_{\tau},\boldsymbol{k}_{\nu} \right) \boldsymbol{A}_{\mathrm{T}}\left( \boldsymbol{l}_{\tau},\boldsymbol{k}_{\nu} \right) $ is approximately the identity matrix $\boldsymbol{I}_P$.
		\end{remark}
%	Different columns in $\boldsymbol{A}_{\mathrm{T}}\left( \boldsymbol{l}_{\tau},\boldsymbol{k}_{\nu} \right) $ can be approximately orthogonal, making $\boldsymbol{A}_{\mathrm{T}}^{\mathrm{H}}\left( \boldsymbol{l}_{\tau},\boldsymbol{k}_{\nu} \right) \boldsymbol{A}_{\mathrm{T}}\left( \boldsymbol{l}_{\tau},\boldsymbol{k}_{\nu} \right) $ close enough to the identity matrix. In addition, the larger the $N$ or $\xi$,
%	the better the orthogonality.
%	the better the orthogonality between different columns in $\boldsymbol{A}_{\mathrm{T}}\left( \boldsymbol{l}_{\tau},\boldsymbol{k}_{\nu} \right) $, which makes $\boldsymbol{A}_{\mathrm{T}}^{\mathrm{H}}\left( \boldsymbol{l}_{\tau},\boldsymbol{k}_{\nu} \right) \boldsymbol{A}_{\mathrm{T}}\left( \boldsymbol{l}_{\tau},\boldsymbol{k}_{\nu} \right) $ closer to the identity matrix.
\end{theorem} 

\begin{IEEEproof}[Proof]
	The proof is given in Appendix B.
\end{IEEEproof}

%The orthogonality between two columns in $\boldsymbol{A}_{\mathrm{T}}\left( \boldsymbol{l}_{\tau},\boldsymbol{k}_{\nu} \right) $ can also be interpreted from the perspective of the distinguishability of the corresponding two paths.
%The two paths in the DAFT domain are sufficiently distinguished so that the correlation between them is low, which makes the orthogonality between the corresponding two columns in $\boldsymbol{A}_{\mathrm{T}}\left( \boldsymbol{l}_{\tau},\boldsymbol{k}_{\nu} \right) $ good enough.
%As $N$ increases, better delay resolution is achieved, which makes the two paths with the same Doppler in the DAFT domain more distinguishable in terms of the delay dimension.
%Similarly, as $\xi$ increases, the two paths with the same delay in the DAFT domain become more distinguishable in the Doppler dimension.

Thus, \eqref{JMLE_lk} and \eqref{h_ML_2} can be well approximated as
%\begin{equation}\label{JMLE_lk_appro}
%	\begin{aligned}
%		\left( \hat{\boldsymbol{l}}_{\tau},\hat{\boldsymbol{k}}_{\tau} \right) \approx& \mathrm{arg}\underset{\left( \boldsymbol{l}_{\tau},\boldsymbol{k}_{\nu} \right)}{\max}\boldsymbol{y}_{\mathrm{T}}^{\mathrm{H}}\boldsymbol{A}_{\mathrm{T}}\left( \boldsymbol{l}_{\tau},\boldsymbol{k}_{\nu} \right) \boldsymbol{A}_{\mathrm{T}}^{\mathrm{H}}\left( \boldsymbol{l}_{\tau},\boldsymbol{k}_{\nu} \right) \boldsymbol{y}_{\mathrm{T}}\\\approx& \mathrm{arg}\max_{\left( \boldsymbol{l}_{\tau},\boldsymbol{k}_{\nu} \right)} \left\| \boldsymbol{A}_{\mathrm{T}}^{\mathrm{H}}\left( \boldsymbol{l}_{\tau},\boldsymbol{k}_{\nu} \right) \boldsymbol{y}_{\mathrm{T}} \right\| ^2\\\approx& \mathrm{arg}\max_{\left( \boldsymbol{l}_{\tau},\boldsymbol{k}_{\nu} \right)} \sum_{i=1}^P{\left| \boldsymbol{a}_{\mathrm{T}}^{\mathrm{H}}\left( l_{\tau _i},k_{\nu _i} \right) \boldsymbol{y}_{\mathrm{T}} \right|^2}.
%	\end{aligned}
%\end{equation}
\begin{equation}\label{JMLE_lk_appro}
	\begin{aligned}
		\left( \hat{\boldsymbol{l}}_{\tau},\hat{\boldsymbol{k}}_{\tau} \right) \approx& \mathrm{arg}\max_{\left( \boldsymbol{l}_{\tau},\boldsymbol{k}_{\nu} \right)} \sum_{i=1}^P{\left| \boldsymbol{a}_{\mathrm{T}}^{\mathrm{H}}\left( l_{\tau _i},k_{\nu _i} \right) \boldsymbol{y}_{\mathrm{T}} \right|^2},
	\end{aligned}
\end{equation}
\begin{equation}
	\hat{\boldsymbol{h}}\approx \small{\frac{\boldsymbol{A}_{\mathrm{T}}^{\mathrm{H}}\left( \hat{\boldsymbol{l}}_{\tau},\hat{\boldsymbol{k}}_{\nu} \right) \boldsymbol{y}_T}{x_p}}.
\end{equation}
Since $\boldsymbol{a}_{\mathrm{T}}\left( l_{\tau _i},k_{\nu _i} \right)$ and $\boldsymbol{a}_{\mathrm{T}}\left( l_{\tau _j},k_{\nu _j} \right)$ are approximately orthogonal, they exhibit path separability, where $i\neq j$.
%In \eqref{JMLE_lk_appro}, we can see that the JMLE of $\left( \hat{\boldsymbol{l}}_{\tau},\hat{\boldsymbol{k}}_{\tau} \right)$ depends on the parameter $\left( l_{\tau _i},k_{\nu _i} \right) $ of the $i$-th path. Based on $\left| \boldsymbol{a}_{\mathrm{T}}^{\mathrm{H}}\left( l_{\tau _i},k_{\nu _i} \right) \boldsymbol{y}_{\mathrm{T}} \right|^2$, $\left( h_i,l_{\tau _i},k_{\nu _i} \right) $ of each path can be estimated separately.
%In addition, Fig. \ref{Separability} illustrates the magnitude of the channel response for two paths with different $N$, where the separability of the channel can also be observed. And 
The separability of the two paths improves as $N$ increases.
In addition, $\boldsymbol{y}_{\mathrm{T}}=\sum_{i=1}^P{x_ph_i\boldsymbol{a}_{\mathrm{T}}\left( l_{\tau _i},k_{\nu _i} \right)}$.
Therefore, $\left( h_i,l_{\tau _i},k_{\nu _i} \right) $ of each path can be estimated separately.
%For $N=256$ and $512$, the delay resolution of the channel is sufficiently high, allowing the two paths to be separated and resulting in two distinct peaks. For $N=128$, the two paths are nearly indistinguishable.
%\begin{figure}[h]
%	\centerline{\includegraphics[scale=0.41 ]{./fig/Separability_of_channels.eps}}
%	\caption{Separability of channels in DAFT domain.}
%	\label{Separability}
%\end{figure}

Therefore, based on the separability of the channel in the DAFT domain, the joint estimation of parameters for all paths can be decoupled. The estimation result for the $i$-th path is
\begin{equation}\label{lk_decoup}
	\left( \hat{l}_{\tau _i},\hat{k}_{\nu _i} \right) \approx \mathrm{arg}\max_{\left( l_{\tau},k_{\nu} \right)} \left| \boldsymbol{a}_{\mathrm{T}}^{\mathrm{H}}\left( l_{\tau},k_{\nu} \right) \boldsymbol{y}_T \right|^2,
\end{equation}
\begin{equation}\label{h_decoup}
	\hat{h}_i\approx \small{\small{\frac{1}{x_p}}\boldsymbol{a}_{\mathrm{T}}^{\mathrm{H}}\left( \hat{l}_{\tau _i},\hat{k}_{\nu _i} \right) \boldsymbol{y}_T}.
\end{equation}
%The estimator in \eqref{lk_decoup} and \eqref{h_decoup} exhibits significantly lower complexity compared to the JMLE in \eqref{JMLE_lk} and \eqref{h_ML_2}. 
%This advantage arises from the elimination of matrix $\boldsymbol{A}_{\mathrm{T}}^{\mathrm{H}}\left( \hat{\boldsymbol{l}}_{\tau},\hat{\boldsymbol{k}}_{\nu} \right) \boldsymbol{A}_{\mathrm{T}}\left( \hat{\boldsymbol{l}}_{\tau},\hat{\boldsymbol{k}}_{\nu} \right) $ inversion and the reduction in the number of parameters to be estimated due to the decoupling of multipath estimation.

\subsection{MF CE for FDFD channels}
For \eqref{lk_decoup}, $\left( \hat{l}_{\tau _i},\hat{k}_{\nu _i} \right) $ can be obtained by searching for consecutive $\left( l_{\tau _i},k_{\nu _i} \right) $.
Specifically, since $\boldsymbol{a}_{\mathrm{T}}\left( l_{\tau _i},k_{\nu _i} \right) $ in $\boldsymbol{y}_{\mathrm{T}}$ and $\boldsymbol{a}_{\mathrm{T}}\left( l_{\tau },k_{\nu} \right) $ have the same structure, it can be anticipated that $\left| \boldsymbol{a}_{\mathrm{T}}^{\mathrm{H}}\left( l_{\tau},k_{\nu} \right) \boldsymbol{y}_T \right|$ achieves its maximum value at $\left( l_{\tau},k_{\nu} \right) =\left( l_{\tau _i},k_{\nu _i} \right) $.
However, the complexity of searching for $\left( \hat{l}_{\tau _i},\hat{k}_{\nu _i} \right) $ in space $\left\{ \left[ 0,l_{\max} \right] \times \left[ -k_{\max},k_{\max} \right] \right\} $ is still high. A simple and effective method is to estimate $\left( \bar{l}_{\tau _i},\bar{k}_{\nu _i} \right) $ of the path based on the maximum peaks in the received signal $\boldsymbol{y}_{\mathrm{T}}$, i.e.,
	\begin{equation}\label{lk_intger_FDFD}
		\left. \hat{\bar{l}}_{\tau _i}=\left. \lfloor \frac{\hat{\chi}_i}{-2Nc_1} \right. \right. \rceil ,\hat{\bar{k}}_{\nu _i}=\hat{\chi}_i+2Nc_1\hat{\bar{l}}_{\tau _i},
\end{equation}
where $\hat{\chi}_i=m_{\max}-n_p$, $m_{\max}=\mathrm{arg}\underset{m}{\max}\left| \left\{ \boldsymbol{y}_{\mathrm{T}} \right\} _m \right|$.

The fractional $\left( \hat{\iota}_{\tau _i},\hat{\kappa}_{\nu _i} \right) $ can be obtained by a global search within a small continuous region $\left\{ \left[ -0.5,0.5 \right] \times \left[ -0.5,0.5 \right] \right\} $. Therefore, $\left( \hat{\iota}_{\tau _i},\hat{\kappa}_{\nu _i} \right) $ can be expressed as
\begin{equation}\label{joint_iota_kappa}
	\left( \hat{\iota}_{\tau _i},\hat{\kappa}_{\nu _i} \right) =\mathrm{arg}\max_{\left( \iota _{\tau},\kappa _{\nu} \right) \in \boldsymbol{\varOmega }} \left| \boldsymbol{a}_{\mathrm{T}}^{\mathrm{H}}\left( \hat{\bar{l}}_{\tau _i}+\iota _{\tau},\hat{\bar{k}}_{\nu _i}+\kappa _{\nu} \right) \boldsymbol{y}_T \right|^2,
\end{equation}
where $\boldsymbol{\varOmega }=\boldsymbol{\varGamma }\times \boldsymbol{\varGamma }\in \mathbb{R} ^{\left( \rho +1 \right) \times \left( \rho +1 \right)}$, $\varGamma _{\tilde{n}}=-0.5+\frac{1}{\rho}\tilde{n}, \tilde{n}\in \left[ 0,\rho \right] $, $\rho$ is the size of the search step.

To further reduce the complexity, a decoupled DD estimation can be considered, i.e.,
\begin{equation}\label{decoupled_iota}
	\hat{\iota}_{\tau _i}=\mathrm{arg}\max_{\iota _{\tau}\in \boldsymbol{\varGamma }} \left| \boldsymbol{a}_{\mathrm{T}}^{\mathrm{H}}\left( \hat{\bar{l}}_{\tau _i}+\iota _{\tau},\hat{\bar{k}}_{\nu _i} \right) \boldsymbol{y}_T \right|^2,
\end{equation}
\begin{equation}\label{decoupled_kappa}
	\hat{\kappa}_{\nu _i}=\mathrm{arg}\max_{\kappa _{\nu}\in \boldsymbol{\varGamma }} \left| \boldsymbol{a}_{\mathrm{T}}^{\mathrm{H}}\left( \hat{\bar{l}}_{\tau _i}+\hat{\iota}_{\tau _i},\hat{\bar{k}}_{\nu _i}+\kappa _{\nu} \right) \boldsymbol{y}_T \right|^2.
\end{equation}
Compared with the joint DD estimation in \eqref{joint_iota_kappa}, which requires $\left( \rho +1 \right) ^2$ searches, the decoupled DD estimation in \eqref{decoupled_iota}–\eqref{decoupled_kappa} involves only $2\left( \rho +1 \right)$ searches. To distinguish between the joint DD estimation and the decoupled DD estimation, they are referred to as MF-JE and MF-DE, respectively.

Noted that $\left( \hat{h}_i,\hat{l}_{\tau _i},\hat{k}_{\nu _i} \right) $ in \eqref{h_decoup}-\eqref{decoupled_kappa} represent the estimation results for single path. For multiple paths, an iterative estimation scheme is required.
%where one path is estimated in each iteration.
Specifically, the signal corresponding to $\left( \hat{l}_{\tau}^{\left( t \right)},\hat{k}_{\nu}^{\left( t \right)},\hat{h}^{\left( t \right)} \right) $ in the $\left(t\right)$-th iteration is
%the path with the maximum peak in $\boldsymbol{y}_{\mathrm{T}}^{\left( t \right)}$ is estimated in $\left(t\right)$-th iteration.
%after obtaining $\left( \hat{l}_{\tau}^{\left( t \right)},\hat{k}_{\nu}^{\left( t \right)},\hat{h}^{\left( t \right)} \right) $ of a path in the $\left(t\right)$-th iteration, the received signal corresponding to this path is
\begin{equation}\label{y_con}
	\boldsymbol{y}_{\mathrm{T}}^{\left( t \right) ,t}=x_p\hat{h}^{\left( t \right)}\boldsymbol{a}\left( \hat{l}_{\tau}^{\left( t \right)},\hat{k}_{\nu}^{\left( t \right)} \right) \in \mathbb{C} ^{\left( Q+1 \right) \times 1}.
\end{equation}
Before estimating the next path, $\boldsymbol{y}_{\mathrm{T}}^{\left( t \right) ,t}$ is subtracted from the received signal $\boldsymbol{y}_{\mathrm{T}}^{\left( t \right)}$, i.e.,
%Once a path estimation is completed, the corresponding received signal $\boldsymbol{y}_{\mathrm{T}}^{\left( t \right) ,t}$ for this path can be subtracted from the received signal $\boldsymbol{y}_{\mathrm{T}}^{\left( t \right)}$, thereby preparing for the estimation of the next path. For $\left( t+1 \right)$-th iteration, we have
\begin{equation}\label{y_t}
	\boldsymbol{y}_{_{\mathrm{T}}}^{\left( t+1 \right)}=\boldsymbol{y}_{_{\mathrm{T}}}^{\left( t \right)}-\boldsymbol{y}_{_{\mathrm{T}}}^{\left( t \right) ,t}.
\end{equation}
The iteration terminates if the maximum number $T_{\mathrm{iter}}$ is reached or $\left| \left\| \boldsymbol{y}_{\mathrm{T}}^{\left( t+1 \right)} \right\| -\left\| \boldsymbol{y}_{\mathrm{T}}^{\left( t \right)} \right\| \right|/\left\| \boldsymbol{y}_{\mathrm{T}}^{\left( t \right)} \right\| \leqslant \sigma $,
%\begin{equation}
%	\left| \small{{{\left\| \boldsymbol{y}_{\mathrm{T}}^{\left( t+1 \right)} \right\| -\left\| \boldsymbol{y}_{\mathrm{T}}^{\left( t \right)} \right\|}\Bigg/{\left\| \boldsymbol{y}_{\mathrm{T}}^{\left( t \right)} \right\|}}} \right|\leqslant \sigma,
%\end{equation}
where the preset threshold $\sigma >0$. The MF with joint DD estimation (MF-JE) CE scheme is summarized in \textbf{Algorithm} \ref{MF_CE_FDD}.

\begin{algorithm}[t]
	\caption{Proposed MF-JE and MF-DE CE scheme for FDFD channels}
	\label{MF_CE_FDD}
	\renewcommand{\algorithmicrequire}{\textbf{Input:}}
	\renewcommand{\algorithmicensure}{\textbf{Output:}}
	\begin{algorithmic}[1]
		\REQUIRE Received signal $\boldsymbol{y}_{\mathrm{T}}$.  %%input
		
		\textbf{Initialization}: $\boldsymbol{y}_{\mathrm{T}}^{\left( t \right)}=\boldsymbol{y}_{\mathrm{T}}$, the maximum number of iterations $T_{\mathrm{iter}}$, threshold $\sigma $, $l_{\max}$, $k_{\max}$, $\hat{\boldsymbol{l}}_{\tau}=\left[  \right] ,\hat{\boldsymbol{k}}_{\nu}=\left[  \right] ,\hat{\boldsymbol{h}}=\left[  \right] $, $t=1$.
		
		\STATE \textbf{repeat}
		\STATE According to \eqref{lk_intger_FDFD}, $\left( \bar{l}_{\tau }^{\left( t \right)},\bar{k}_{\nu }^{\left( t \right)} \right) $ is obtained.
		\STATE $\left( \hat{\iota}_{\tau }^{\left( t \right)},\hat{\kappa}_{\nu }^{\left( t \right)} \right) $ is estimated via \eqref{joint_iota_kappa} in MF-JE and via \eqref{decoupled_iota}–\eqref{decoupled_kappa} in MF-DE.
		\STATE $\hat{h}^{\left( t \right)}$ can be obtained by \eqref{h_decoup}.
		\STATE $\hat{\boldsymbol{l}}_{\tau}=\left[ \hat{\boldsymbol{l}}_{\tau},\hat{l}_{\tau }^{\left( t \right)} \right] $, $\hat{\boldsymbol{k}}_{\nu}=\left[ \hat{\boldsymbol{k}}_{\nu},\hat{k}_{\nu }^{\left( t \right)} \right] $, $\hat{\boldsymbol{h}}=\left[ \hat{\boldsymbol{h}},\hat{h}^{\left( t \right)} \right] $.
		\STATE The $\boldsymbol{y}_{\mathrm{T}}^{\left( t+1 \right)}$ is obtained by \eqref{y_t}.
		\STATE $t=t+1$.
		\STATE until $t= T_{\mathrm{iter}}$ or $\left| \small{\frac{\left\| \boldsymbol{y}_{\mathrm{T}}^{\left( t+1 \right)} \right\| -\left\| \boldsymbol{y}_{\mathrm{T}}^{\left( t \right)} \right\|}{\left\| \boldsymbol{y}_{\mathrm{T}}^{\left( t \right)} \right\|}} \right|\leqslant \sigma$.
		
		\ENSURE $\left( \hat{\boldsymbol{l}}_{\tau},\hat{\boldsymbol{k}}_{\nu},\hat{\boldsymbol{h}} \right) $.    %%output
	\end{algorithmic}
\end{algorithm}

%Note that the relative positions of the peaks of the responses of the $P$ paths in the received signals corresponding to $\boldsymbol{x}_1$ and $\boldsymbol{x}_2$ do not change.

\subsection{MF-GFS-DE CE for FDFD channels}
In Subsection \Rmnum{4}-B, an MF CE scheme is proposed.
%This scheme employs an iterative approach to sequentially estimate the delay, Doppler, and channel gain for each propagation path. 
%This scheme eliminates the matrix inversion operation, decouples joint estimation, and narrows the search range to significantly reduce the computational complexity. 
However, the trade-off between the complexity and estimation performance of the MF CE scheme yet depends on the $\rho$. Performing a global search over an interval is inefficient, and the estimation accuracy is limited by the search step size $\rho$.
	%GFS is an unconstrained nonlinear optimization method for unimodal functions \cite{Avriel,Subasi,Chong}.
	%It is the most effective derivative-free method for minimizing strictly unimodal functions in a closed bounded interval.
The larger the $\rho$, the finer the grid $\boldsymbol{\varGamma }$, enabling more accurate estimation of fractional part at the cost of higher complexity.
%In addition, the received signal corresponding to a certain path reconstructed by \eqref{y_con} is also more accurate. 
%In this case, $\left( \hat{\boldsymbol{l}}_{\tau},\hat{\boldsymbol{k}}_{\nu},\hat{\boldsymbol{h}} \right) $ will become more accurate. 
%However, the cost of increasing $\rho$ is that the complexity will inevitably increase.
To address this issue, we introduce a MF-GFS CE scheme based on the GFS algorithm in this subsection. The GFS algorithm unevenly divides the search interval into two parts through the ratio of two consecutive generalized Fibonacci numbers (GFNs) to achieve an efficient search.
The $i$-th GFN is denoted by $S_i\left( a,b,p,q \right) $, where $S_0=a$, $S_1=b$, $S_{i+2}=pS_{i+1}+qS_{i}$, $i$ is a non-negative integer, $a$ and $b$ be two non-negative integers such that $a+b>0$, $p$ and $q$ be two positive integers.

%The $\left(i+2\right)$-th GFN is a linear combination of its two preceding GFNs, denoted as $S_{i+2}=pS_{i+1}+qS_i$, $S_0=a$, $S_1=b$. The Fibonacci numbers, with $p=q=1$, $S_0=0$, $S_1=1$, and the Lucas numbers, with $p=q=1$, $S_0=2$, $S_1=1$, are specific cases of the GFNs.

Note that an important pre-condition for using the GFS algorithm is that the objective function should be a bounded unimodal function.
%The first condition is easy to satisfy. 
%In addition, the objective function in \eqref{chi_f} needs to satisfy the second condition.
Since fractional delays may cause local peaks, the objective functions in both \eqref{joint_iota_kappa} and \eqref{decoupled_iota} are generally not unimodal.
\newtheorem{theorem3}{\bf Theorem}
\begin{theorem}\label{thm3}
	In \eqref{decoupled_kappa}, $\left| \boldsymbol{a}_{\mathrm{T}}^{\mathrm{H}}\left( \hat{\bar{l}}_{\tau _i}+\hat{\iota}_{\tau _i},\hat{\bar{k}}_{\nu _i}+\kappa _{\nu} \right) \boldsymbol{y}_T \right|^2$ is a bounded unimodal function.
\end{theorem} 
\begin{IEEEproof}[Proof]
	The proof is given in Appendix C.
\end{IEEEproof}

%For the GFS algorithm, it is feasible to start partitioning the search interval using the ratio of two consecutive larger or smaller GFNs. There is no difference in the results of the two methods.
%However, if the ratio of larger GFNs is considered, the response at only one point needs to be calculated in each iteration rather than two. 
%This paper considers dividing the search interval starting from the ratio of two consecutive large GFNs to offer lower complexity. To achieve this, $p=q=1$ must be satisfied.
After dividing the search interval by two consecutive GFNs, the interval of uncertainty (IU) containing the optimal value $\kappa _{\nu_i}$ needs to be determined. For convenience, assume that the search interval is $\left[ x_s,x_f \right] $, and $x_s<x_1<x_2<x_f$.
%\begin{equation}
%	x_s<x_1<x_2<x_f.
%\end{equation}
%Finding the IU will involve calculating and comparing the corresponding function values at different points in the $\left[ x_s,x_f \right] $ interval.
The new IU $\left[ x_{s}^{\prime},x_{f}^{\prime} \right]$ can be expressed as
\begin{equation}\label{new_IU}
	\left[ x_{s}^{\prime},x_{f}^{\prime} \right] =\left\{ \begin{matrix}	\left[ x_s,x_2 \right] ,&		f\left( x_1 \right) <f\left( x_2 \right)\\	\left[ x_1,x_f \right] ,&		f\left( x_1 \right) >f\left( x_2 \right)\\\end{matrix} \right. .
\end{equation}
%\begin{itemize}
%	\item{If $f\left( x_1 \right) <f\left( x_2 \right) $,
%%		there will be two possibilities, i.e., $x_s<\kappa _i<x_1<x_2<x_f$ or $x_s<x_1<\kappa _i<x_2<x_f$. Therefore, 
%		the new IU can be expressed as
%		\begin{equation}\label{IU1}
%			\left[ x_{s}^{\prime},x_{f}^{\prime} \right] =\left[ x_s,x_1 \right] \bigcup{\left[ x_1,x_2 \right]}=\left[ x_s,x_2 \right],
%		\end{equation}
%		where $\hat{\gamma} _i\in \left( x_{s}^{\prime},x_{f}^{\prime} \right) $, $x_{s}^{\prime}=x_s$, $x_{f}^{\prime}=x_2$.}
%	
%	\item{If $f\left( x_1 \right) >f\left( x_2 \right) $,
%%		there are also two possibilities, i.e., $x_s<x_1<\kappa _i<x_2<x_f$ or $x_s<x_1<x_2<\kappa _i<x_f$. Similarly, 
%		the new IU is
%		\begin{equation}\label{IU2}
%			\left[ x_{s}^{\prime},x_{f}^{\prime} \right] =\left[ x_1,x_2 \right] \bigcup{\left[ x_2,x_f \right]}=\left[ x_1,x_f \right] ,
%		\end{equation}
%		where $\hat{\gamma} _i\in \left( x_{s}^{\prime},x_{f}^{\prime} \right) $, $x_{s}^{\prime}=x_1$, $x_{f}^{\prime}=x_f$.}
%\end{itemize}
%Note that $\left| x_{f}^{\prime}-x_{s}^{\prime} \right|<\left| x_f-x_s \right|$.

\begin{figure}[t]
	\centerline{\includegraphics[width=\columnwidth ]{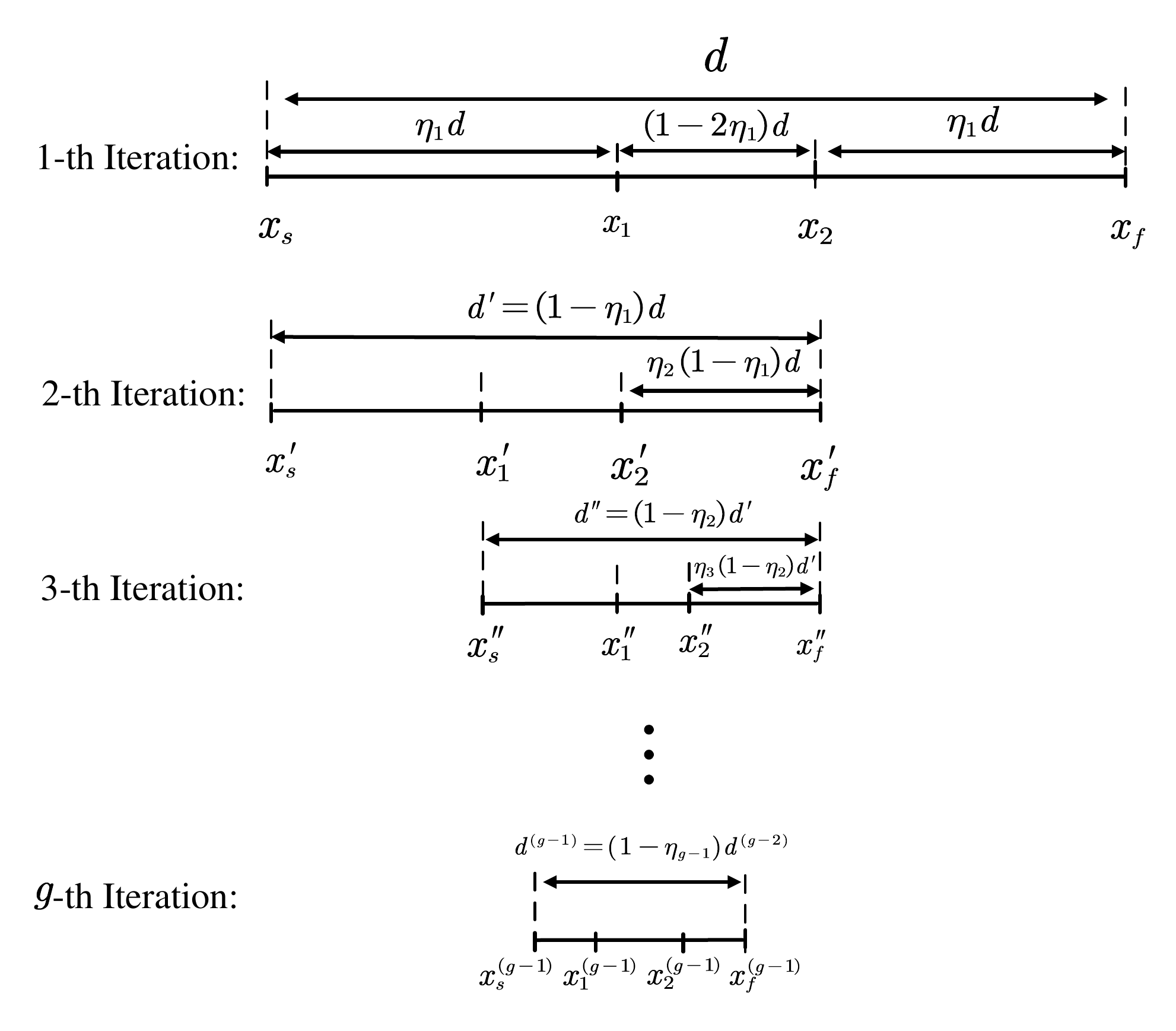}}
	\caption{The diagram of the GFS algorithm.}
	\label{GFS}
\end{figure}

The process of the GFS for estimating $\hat{\kappa} _{\nu_i}$ is shown in Fig. \ref{GFS}.
Note that it is necessary to set a maximum GFN $S_{N_G} $, $N_G=T_G+2$, which determines the maximum number of iterations, $T_G$, for the GFS. For the $g$-th iteration, the IU is divided based on the ratio of two successive higher GFNs, i.e.,
\begin{equation}\label{IU_gamma}
	\eta _g=\small{\frac{qS_{N_G-\left( g+1 \right)}\left( a,b,p,q \right)}{S_{N_G-\left( g-1 \right)}\left( a,b,p,q \right)}},
\end{equation}
where $g\in \left[ 1,T_G \right] $. $\eta _g$ can determine the positions of $x_1^{\left(g-1\right)}$ and $x_2^{\left(g-1\right)}$.
%Specifically, assume that $\left| x_f-x_s \right|=d$, $\left| x_1-x_s \right|=\left| x_f-x_2 \right|$. For the first iteration, let $\frac{x_1-x_s}{d}=\frac{x_f-x_2}{d}=\eta _1$, where $\eta _1=\frac{qS_{N_G-2}\left( a,b,p,q \right)}{S_{N_G}\left( a,b,p,q \right)}$ is used to determine the positions of $x_1$ and $x_2$. A new IU $\left[ x_{s}^{\prime},x_{f}^{\prime} \right] $ will be obtained by \eqref{IU1} or \eqref{IU2}. In the second iteration, the size of IU obtained by the first iteration is $d\prime=x_{f}^{\prime}-x_{s}^{\prime}=\left( 1-\eta _1 \right) d$. Similarly, let $\frac{x_{1}^{\prime}-x_{s}^{\prime}}{d\prime}=\frac{x_{f}^{\prime}-x_{2}^{\prime}}{d\prime}=\eta _2$ to determine the positions of $x_{1}^{\prime}$ and $x_{2}^{\prime}$. 
%Since $x_{f}^{\prime}-x_{2}^{\prime}=x_2-x_1=\left( 1-2\eta _1 \right) d$, $\eta _2\left( 1-\eta _1 \right) d=\left( 1-2\eta _1 \right) d$, which leads to $\eta _2=\small{\frac{qS_{N_G-3}\left( a,b,p,q \right)}{S_{N_G-1}\left( a,b,p,q \right)}}$. Then, repeat the previous process. 
In general, for the $g$-th iteration, the size of IU is
\begin{equation}\label{IU_d}
	d^{\left( g-1 \right)}=\left( 1-\eta _{g-1} \right) d^{\left( g-2 \right)}.
\end{equation}
Then, the new IU can be obtained by \eqref{new_IU}.
The iteration will terminate if the maximum number $T_G$ is reached or
\begin{equation}
	d^{\left( g-1 \right)}=\left| x_{f}^{\left( g-1 \right)}-x_{s}^{\left( g-1 \right)} \right|<\,\,\varepsilon,
\end{equation}
where $\varepsilon$ is the pre-set threshold. At this moment, 
%the size $d^{\left( g \right)}$ of IU will be very small.
if there is no additional prior, fractional part can be estimated as
\begin{equation}\label{IU_kappa}
	\hat{\kappa}_{\nu_i}=\frac{{x}_s^{\left( g-1 \right)}+{x}_f^{\left( g-1 \right)}}{2}.
\end{equation}
%where $\tilde{x}_s$ and $\tilde{x}_f$ are the starting and final positions of IU when the iteration is terminated.
The proposed GFS algorithm for estimating $\hat{\kappa}_{\nu_i}$ is summarized in \textbf{Algorithm} \ref{GFS Algorithm}.
Except for $\hat{\kappa}_{\nu_i}$, the MF-GFS scheme follows the same procedure as the MF scheme. For the MF-GFS scheme, $\hat{\kappa}_{\nu_i}$ is obtained via the \textbf{Algorithm} \ref{GFS Algorithm} rather than \eqref{decoupled_kappa}.

Note that the size of IU in the global search method is $\small{\frac{1}{\rho +1}}$, which is generally on the magnitude of $10^{-1}$ or $10^{-2}$. However, the size of IU in the GFS is several orders of magnitude smaller. Therefore, the proposed MF-GFS scheme is generally more effective than the MF scheme.

\begin{algorithm}[t]
	\caption{The proposed GFS Algorithm}
	\label{GFS Algorithm}
	\renewcommand{\algorithmicrequire}{\textbf{Input:}}
	\renewcommand{\algorithmicensure}{\textbf{Output:}}
	\begin{algorithmic}[1]
		\REQUIRE Received signal $\boldsymbol{y}_{\mathrm{T}}$, the objective function $f\left( \kappa_{\nu} \right) =\left| \boldsymbol{a}_{\mathrm{T}}^{\mathrm{H}}\left( \hat{\bar{l}}_{\tau _i}+\hat{\iota}_{\tau _i},\hat{\bar{k}}_{\nu _i}+\kappa _{\nu} \right) \boldsymbol{y}_T \right|^2$.
		
		\textbf{Initialization}: $g=1$, $T_G$, IU $\left[ x_{s}^{\left( 0 \right)},x_{f}^{\left( 0 \right)} \right] =\left[ -0.5,0.5 \right] $.
		\STATE \textbf{repeat}
		\STATE From \eqref{IU_d} and \eqref{IU_gamma}, $x_{1}^{\left( g-1 \right)}$ and $x_{2}^{\left( g-1 \right)}$ are obtained.
		\STATE According to \eqref{new_IU}, the new IU $\left[ x_{s}^{\left( g \right)},x_{f}^{\left( g \right)} \right]$ is obtained.
		\STATE $g=g+1$.
		\STATE \textbf{until} $g= T_G$ or $d^{\left( g-1 \right)}=\left| x_{f}^{\left( g-1 \right)}-x_{s}^{\left( g-1 \right)} \right|<\,\,\varepsilon $.
		\STATE Calculate $\hat{\kappa}_{\nu_i}$ according to \eqref{IU_kappa}.
		
		\ENSURE $\hat{\kappa}_{\nu_i}$.    %%output
	\end{algorithmic}
\end{algorithm}

\begin{algorithm}[t]
	\caption{The proposed MF and MF-GFS CE scheme for IDFD channels}
	\label{MF_GFS_CE}
	\renewcommand{\algorithmicrequire}{\textbf{Input:}}
	\renewcommand{\algorithmicensure}{\textbf{Output:}}
	\begin{algorithmic}[1]
		\REQUIRE Received signal $\boldsymbol{y}_{\mathrm{T}}$, the objective function $\left| \boldsymbol{a }_{\mathrm{T}}^{\mathrm{H}}\left( \hat{\bar{l}}_{\tau _i},\hat{\bar{k}}_{\nu _i}+\kappa _{\nu} \right) \boldsymbol{y}_T \right|^2$.  %%input
		
		\textbf{Initialization}: $\boldsymbol{y}_{\mathrm{T}}^{\left( t \right)}=\boldsymbol{y}_{\mathrm{T}}$, the maximum number of iterations $T_{\mathrm{iter}}$ and $T_G$, threshold $\sigma $ and $\epsilon$, $l_{\max}$, $k_{\max}$, $\hat{\boldsymbol{l}}_{\tau}=\left[  \right] ,\hat{\boldsymbol{k}}_{\nu}=\left[  \right] ,\hat{\boldsymbol{h}}=\left[  \right] $, $t=1$.
		
		\STATE \textbf{repeat}
		\STATE According to \eqref{lk_intger_FDFD}, $\left(\hat{\bar{l}}_{\tau}^{\left( t \right)},\hat{\bar{k}}_{\nu}^{\left( t \right)} \right) $ for a path is obtained.
		\STATE $\hat{\kappa}_{\nu }^{\left( t \right)}$ is estimated via \eqref{kappa} in MF and via \textbf{Algorithm} \ref{GFS Algorithm} in MF-GFS.
		\STATE $\hat{h}^{\left( t \right)}$ can be obtained by \eqref{h_decoup}.
		\STATE $\hat{\boldsymbol{l}}_{\tau}=\left[ \hat{\boldsymbol{l}}_{\tau},\hat{l}_{\tau}^{\left( t \right)} \right] $, $\hat{\boldsymbol{k}}_{\nu}=\left[ \hat{\boldsymbol{k}}_{\nu},\hat{k}_{\nu}^{\left( t \right)} \right] $, $\hat{\boldsymbol{h}}=\left[ \hat{\boldsymbol{h}},\hat{h}^{\left( t \right)} \right] $.
		\STATE $\boldsymbol{y}_{\mathrm{T}}^{\left( t+1 \right)}$ is obtained by \eqref{y_t}.
		\STATE $t=t+1$.
		\STATE until $t= T_{\mathrm{iter}}$ or $\left| \small{\frac{\left\| \boldsymbol{y}_{\mathrm{T}}^{\left( t+1 \right)} \right\| -\left\| \boldsymbol{y}_{\mathrm{T}}^{\left( t \right)} \right\|}{\left\| \boldsymbol{y}_{\mathrm{T}}^{\left( t \right)} \right\|}} \right|\leqslant \sigma$.
		
		\ENSURE $\left( \hat{\boldsymbol{l}}_{\tau},\hat{\boldsymbol{k}}_{\nu},\hat{\boldsymbol{h}} \right) $.    %%output
	\end{algorithmic}
\end{algorithm}

\subsection{MF and MF-GFS CE schemes for IDFD channels}
In typical wide-band systems such as satellite or millimeter-wave communications, $B$ is sufficiently large to approximate $\tau_i$ to the nearest sampling point \cite{P. Raviteja_1}.
Therefore, MF and MF-GFS schemes are proposed for IDFD channels.

The observation model in \eqref{y_T} is still considered. The estimation of $\left( h_i,l_{\tau _i}=\bar{l}_{\tau _i},k_{\nu _i}=\bar{k}_{\nu _i}+\kappa _{\nu _i} \right) $ for the $i$-th path can also be expressed as \eqref{lk_decoup} and \eqref{h_decoup}.

To reduce complexity, the integer $\left( \bar{l}_{\tau _i},\bar{k}_{\nu _i} \right) $ is initially also estimated using \eqref{lk_intger_FDFD}.
$\hat{\kappa}_{\nu _i}$ can be expressed as
\begin{equation}\label{kappa}
	\hat{\kappa}_{\nu _i}=\mathrm{arg}\max_{\kappa _{\nu}\in \boldsymbol{\varGamma }} \left| \boldsymbol{a }_{\mathrm{T}}^{\mathrm{H}}\left( \hat{\bar{l}}_{\tau _i},\hat{\bar{k}}_{\nu _i}+\kappa _{\nu} \right) \boldsymbol{y}_T \right|^2.
\end{equation}

The MF scheme is to directly search the grid $\boldsymbol{\varGamma }$ to get $\hat{\kappa}_{\nu _i}$.
In addition, $\left| \boldsymbol{a }_{\mathrm{T}}^{\mathrm{H}}\left( \hat{\bar{l}}_{\tau _i},\hat{\bar{k}}_{\nu _i}+\kappa _{\nu} \right) \boldsymbol{y}_T \right|^2$ is a bounded unimodal function, achieving its extreme value at $\kappa _{\nu}=\kappa _{\nu _i}$. This proof is similar to Appendix C.
Therefore, to reduce complexity and improve performance, \textbf{Algorithm} \ref{GFS Algorithm} is used to estimate $\hat{\kappa}_{\nu _i}$. 
Note that the same iterative method is also considered for parameter estimation of multiple paths, which estimates one path in one iteration.
The proposed  MF and MF-GFS CE schemes for IDFD channels are summarized in \textbf{Algorithm} \ref{MF_GFS_CE}.

%\begin{figure*}[t] % hb底部，ht为头部
%	\centering % 公式居中
%	\hrulefill % 添加一条水平线
%	%	\vspace*{8pt} % 调整线与公式之间的距离
%	%\begin{equation}
%	\begin{align}
%		\left| \boldsymbol{a }_{\mathrm{T}}^{\mathrm{H}}\left( \hat{\bar{l}}_{\tau _i},\hat{\bar{k}}_{\nu _i}+\kappa _{\nu} \right) \boldsymbol{y}_T \right|
%		=&\left| \boldsymbol{a}_{\mathrm{T}}^{\mathrm{H}}\left( \hat{\bar{l}}_{\tau _i},\hat{\bar{k}}_{\nu _i}+\kappa _{\nu} \right) \sum_{j=1}^P{x_ph_i\boldsymbol{a}_{\mathrm{T}}\left( l_{\tau _j},k_{\nu _j} \right)} \right|\label{CC-1}
%		\\\approx& \left| x_ph_i\boldsymbol{a}_{\mathrm{T}}^{\mathrm{H}}\left( \hat{\bar{l}}_{\tau _i},\hat{\bar{k}}_{\nu _i}+\kappa _{\nu} \right) \boldsymbol{a}_{\mathrm{T}}\left( l_{\tau _i},k_{\nu _i} \right) \right|\label{CC-2}
%		\\\approx& \left| \frac{1}{N}x_ph_i\frac{\sin \left\{ \pi \left[ 2Nc_1(l_{\tau _i}-\hat{\bar{l}}_{\tau _i})-\left( \left( \bar{k}_{\nu _i}+\kappa _{\nu _i} \right) -\left( \hat{\bar{k}}_{\nu _i}+\kappa _{\nu} \right) \right) \right] \right\}}{\sin \left\{ \frac{\pi}{N}\left[ 2Nc_1(l_{\tau _i}-\hat{\bar{l}}_{\tau _i})-\left( \left( \bar{k}_{\nu _i}+\kappa _{\nu _i} \right) -\left( \hat{\bar{k}}_{\nu _i}+\kappa _{\nu} \right) \right) \right] \right\}} \right|\label{CC-3}
%		\\\approx& \left| \frac{1}{N}x_ph_i\frac{\sin \left\{ \pi \left( \kappa _{\nu}-\kappa _{\nu _i} \right) \right\}}{\sin \left\{ \frac{\pi}{N}\left( \kappa _{\nu}-\kappa _{\nu _i} \right) \right\}} \right|\label{CC-4}
%		\\\approx& \left| x_ph_i \right|\mathrm{sinc}\left( \kappa _{\nu}-\kappa _{\nu _i} \right)\label{CC-5}
%	\end{align}
%	%\end{equation}
%\end{figure*}

\subsection{Complexity Analysis }
\begin{figure}[t]
	\centerline{\includegraphics[width=\columnwidth ]{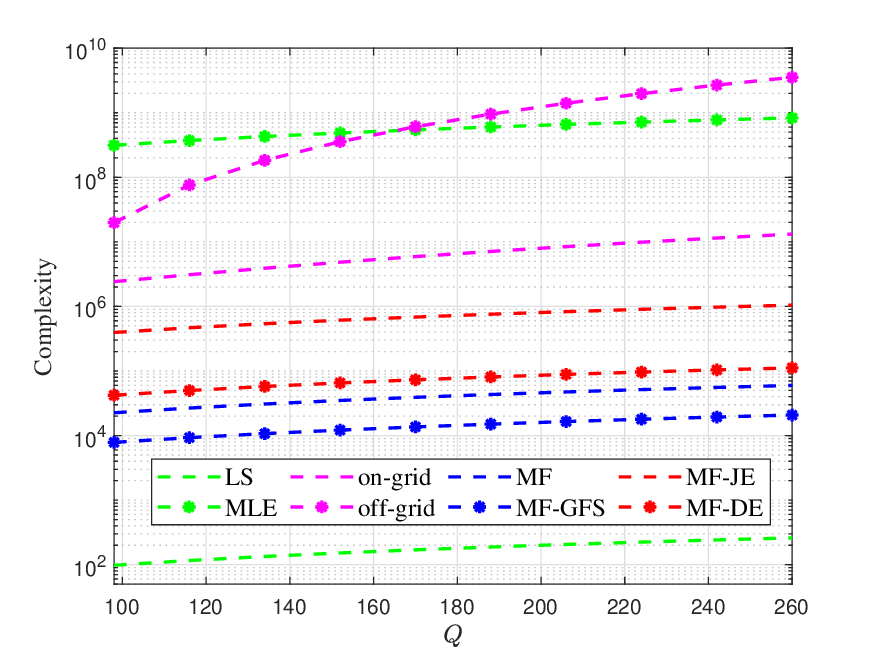}}
	\caption{Complexity comparison of different CE schemes.}
	\label{complexty}
\end{figure}

In this subsection, we analyze complexity of the proposed MF and MF-GFS CE schemes. For the MF CE in the IDFD channels, 
The complexity of \eqref{lk_intger_FDFD} is $\mathcal{O} \left(Q\right)$.
In addition, based on \eqref{kappa}, estimating $\hat{\kappa}_{\nu _i}$ requires $\mathcal{O} \left( Q\rho \right) $.
Calculating \eqref{h_decoup} costs $\mathcal{O} \left(Q\right)$. To calculate $\boldsymbol{y}_{\mathrm{T}}^{\left( t+1 \right) }$ in \eqref{y_t}, $\mathcal{O} \left( Q \right)$ are required. Therefore, the complexity in each iteration is $\mathcal{O} \left( Q\rho+3Q \right) $. For the proposed MF-GFS CE scheme in the IDFD channel, 
estimating $\hat{\kappa}_{\nu _i}$ through \textbf{Algorithm} \ref{GFS Algorithm} costs $\mathcal{O} \left(T_GQ\right)$. Therefore, the complexity in each iteration is $\mathcal{O} \left( T_GQ+3Q \right)$.
The remaining analysis is similar to the MF scheme.
The complexity is summarized in Table \ref{complexity}, where $L$ is the number of columns in the measurement matrix for the CS schemes.
Fig. \ref{complexty} is the complexity comparison, where $l_{\max}=8$, $k_{\max} \in \left[2,10\right]$. one can see the low-complexity nature of the proposed schemes.

\begin{table*}[t]\caption{\textsc{Comparison of Complexity}}
	\centering
	\label{complexity}
	\begin{tabular}{|c|c|}
		\hline
		Method   & Complexity   \\\hline
		LS scheme \cite{H. Yin_1} & $\mathcal{O}\left(Q\right)$\\\hline
		MLE scheme\cite{Bemani_3} & $\mathcal{O} \left( \rho ^PQ \right) $\\\hline
		on-grid scheme \cite{Benzine_1,Benzine_2} & $\mathcal{O} \left( T_{\mathrm{iter}}^{3}+T_{\mathrm{iter}}LQ+T_{\mathrm{iter}}^{2}Q \right) $\\\hline
		off-grid scheme \cite{F. Yang} & $\mathcal{O} \left( T_{\mathrm{iter}}L^2Q \right) $\\\hline
		The proposed MF scheme for IDFD channels& $\mathcal{O}\left(T_{\mathrm{iter}}\left(\rho Q +3Q\right)\right)$\\\hline
		The proposed MF-GFS scheme for IDFD channels& $\mathcal{O}\left( T_{\mathrm{iter}}\left(T_GQ +3Q\right) \right) $\\\hline
		The proposed MF-JE and MF-DE scheme for FDFD channels& $\mathcal{O}\left(T_{\mathrm{iter}}\left(\rho^2 Q +3Q\right) \right)$, $\mathcal{O} \left( T_{\mathrm{iter}}\left( 2\rho Q+3Q \right) \right) $\\\hline
		The proposed MF-GFS-DE scheme for FDFD channels& $\mathcal{O} \left( T_{\mathrm{iter}}\left( \rho +T_G \right) Q+3Q \right) $\\\hline
	\end{tabular}
\end{table*}

\section{Simulation Results}
\begin{table}[t]\caption{\textsc{Simulation Parameters}}
	\centering
	\label{parameters}
	\begin{tabular}{|c|c|}
		\hline
		Parameter   & Value   \\\hline
		The number of chirp-carriers& $N=256$\\\hline
		Carrier frequency& $4\text{ GHz}$\\\hline
		The channel coefficient  & $h_i\sim \mathcal{C} \mathcal{N} \left( 0,\frac{1}{P} \right) $ \\\hline
		Maximum delay & ${{\tau }_{\max }}=1.56\times {{10}^{-5}}\text{s}$ \\\hline
		Maximum relative velocity   &  $540\text{ km/h}$\\\hline
		Maximum Doppler shift   &  $2\times {{10}^{3}}$ Hz\\\hline
		Subcarrier spacing & $\Delta f\text{ }=\text{ }1\text{ kHz}$\\\hline
		Number of channel paths   & $P\text{ }=\text{ }5$ \\\hline
		Data modulation   & $4$-QAM \\\hline
		$\xi$, $\xi^{\prime}$   & $4$, $5$ \\\hline
	\end{tabular}
\end{table}
In this section, the CE performances of the proposed MF and MF-GFS methods are evaluated. Unless otherwise stated, the simulation parameters are shown in Table \ref{parameters}.
%The maximum Doppler tap and delay tap corresponding to the maximum Doppler shift and maximum delay are given by $k_{\max}=2$ and $l_{\max}=4$, respectively.
%According to the maximum relative velocity and $\tau_{\max}$, $k_{\max}=2$ and $l_{\max}=4$.
$\tau _i$ and $\nu _i$ are randomly and uniformly generated within $\left[ 0,\tau _{\max} \right] $ and $\left[ -\nu _{\max},\nu _{\max} \right] $, respectively. The system signal-to-noise ratio (SNR) is defined as $\frac{E\left\{ \left| x\left[ n \right] \right|^2 \right\}}{\sigma ^2}$.
For the pilot $x_p$, $10\log _{10}\frac{\left| x_p \right|^2}{E\left\{ \left| x\left[ n \right] \right|^2 \right\}}=30$ dB is satisfied.
%The power of the pilot symbol $x_p$ is set to be $30$ dB higher than that of the data symbols, i.e., $10\log _{10}\frac{\left| x_p \right|^2}{E\left\{ \left| x\left( n \right) \right|^2 \right\}}=30\mathrm{dB}$.
For the proposed algorithms, set $T_{\mathrm{iter}}=15$, $\sigma =10^{-3}$, $\rho=20$. For comparison, we consider the traditional LS scheme \cite{H. Yin_1}, the MLE scheme \cite{Bemani_3}, the on-grid scheme \cite{Benzine_1,Benzine_2} with OMP algorithm, and the off-grid scheme \cite{F. Yang}. For CE schemes based on CS algorithm, the grid resolution is 0.5.
We reconstruct $\hat{\boldsymbol{H}}_{\mathrm{eff}}$ using the estimated parameters, and define NMSE as $\mathrm{NMSE}\left. = \right\| \boldsymbol{H}_{\mathrm{eff}}-\hat{\boldsymbol{H}}_{\mathrm{eff}}\left\| ^2 \right. /\left\| \boldsymbol{H}_{\mathrm{eff}} \right\| ^2$.

	\begin{figure}[t]
		\centerline{\includegraphics[width=\columnwidth ]{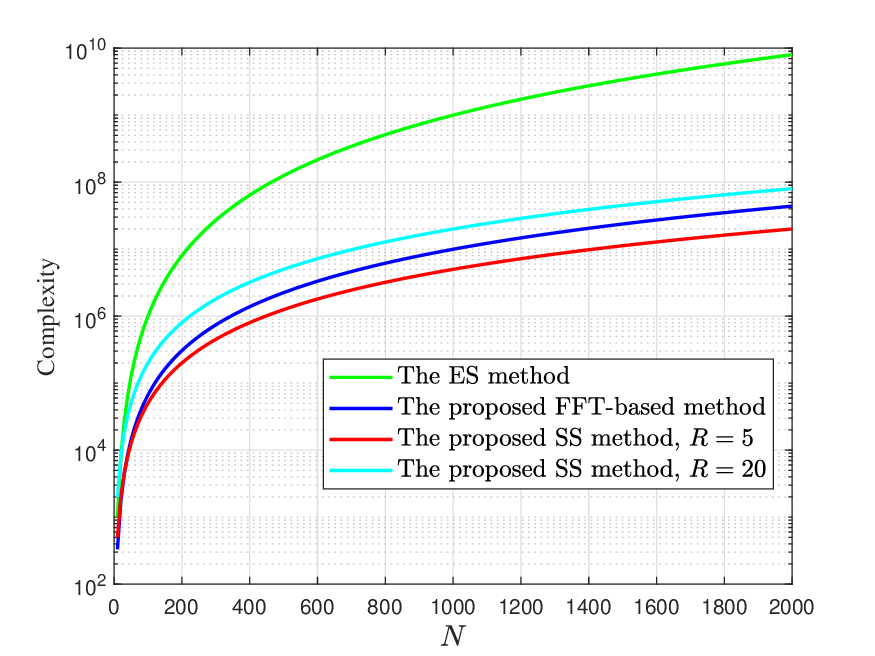}}
		\caption{Complexity of constructing $\boldsymbol{H}_{\mathrm{eff}}$ via different methods.}
		\label{Channel_matrix_complexity}
	\end{figure}
	Fig. \ref{Channel_matrix_complexity} shows a comparison of the complexity of constructing $\boldsymbol{H}_{\mathrm{eff}}$ using different methods. For large $N$, the proposed FFT-based and SS methods reduce complexity by at least two orders of magnitude. The SS method is preferable for small $R$, while the FFT-based method is better for large $R$.

\begin{figure}[t]
	\centerline{\includegraphics[width=\columnwidth ]{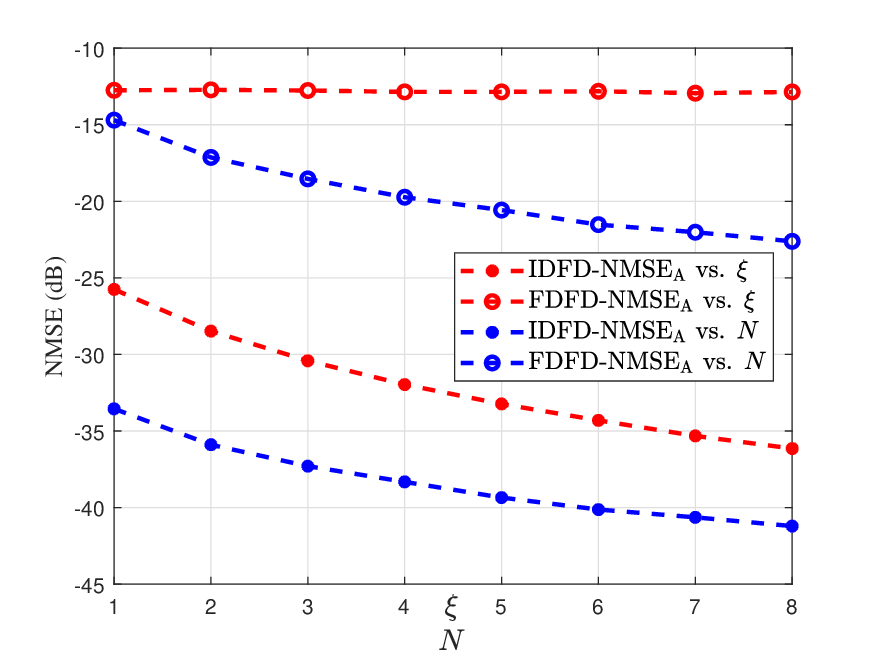}}
%	\caption{Approximation error of $\boldsymbol{A}_{\mathrm{T}}^{\mathrm{H}}\left( \boldsymbol{l}_{\tau},\boldsymbol{k}_{\nu} \right) \boldsymbol{A}_{\mathrm{T}}\left( \boldsymbol{l}_{\tau},\boldsymbol{k}_{\nu} \right) $ approximation to identity matrix $\boldsymbol{I}$ vs. $\xi$ and $N$.}
	\caption{Orthogonality of $\boldsymbol{A}_{\mathrm{T}}\left( \boldsymbol{l}_{\tau},\boldsymbol{k}_{\nu} \right) $ vs. $\xi$ and $N$.}
	\label{IM_xi_N}
\end{figure}
The approximate orthogonality of $\boldsymbol{A}_{\mathrm{T}}\left( \boldsymbol{l}_{\tau},\boldsymbol{k}_{\nu} \right) $ for different values of $\xi$ and $N$ is shown in Fig. \eqref{IM_xi_N}.
For “$\text{NMSE}_\text{A}$ vs. $\xi$”, we consider $ N=256$, $\xi \in \left[ 1,8 \right] $. For “$\text{NMSE}_\text{A}$ vs. $N$”, we have $ \xi=4$, $N\in 200\times \left[ 1,8 \right] $.
To make this orthogonality more obvious, the integer parts of the delays for different paths are set to differ. Note that $\text{NMSE}_\text{A}\left. = \right\| \boldsymbol{I}-\boldsymbol{A}_{\mathrm{T}}^{\mathrm{H}}\left( \boldsymbol{l}_{\tau},\boldsymbol{k}_{\nu} \right) \boldsymbol{A}_{\mathrm{T}}\left( \boldsymbol{l}_{\tau},\boldsymbol{k}_{\nu} \right) \left\| ^2 \right. /\left\| \boldsymbol{I} \right\| ^2$.
%\begin{equation}
%	\mathrm{NMSE}_{\mathrm{A}}=\frac{\left\| \boldsymbol{I}-\boldsymbol{A}_{\mathrm{T}}^{\mathrm{H}}\left( \boldsymbol{l}_{\tau},\boldsymbol{k}_{\nu} \right) \boldsymbol{A}_{\mathrm{T}}\left( \boldsymbol{l}_{\tau},\boldsymbol{k}_{\nu} \right) \right\| ^2}{\left\| \boldsymbol{I} \right\| ^2}.
%\end{equation}
It can be observed that this approximate orthogonality is still good enough even in fractional channels.
%$\boldsymbol{A}_{\mathrm{T}}^{\mathrm{H}}\left( \boldsymbol{l}_{\tau},\boldsymbol{k}_{\nu} \right) \boldsymbol{A}_{\mathrm{T}}\left( \boldsymbol{l}_{\tau},\boldsymbol{k}_{\nu} \right) $ can still be effectively approximated as the identity matrix, even in fractional channels.
This orthogonality improves with the increase of $\xi$ or $N$. However, for FDFD channels, increasing $\xi$ does not improve orthogonality.
%The fact that $\boldsymbol{A}_{\mathrm{T}}^{\mathrm{H}}\left( \boldsymbol{l}_{\tau},\boldsymbol{k}_{\nu} \right) \boldsymbol{A}_{\mathrm{T}}\left( \boldsymbol{l}_{\tau},\boldsymbol{k}_{\nu} \right) $ can approximate the identity matrix well can also be seen from Appendix B.

%\begin{figure}[h]
%	\centerline{\includegraphics[width = 7cm ]{./fig/simulation/NMSE_phase.eps}}
%	\caption{Effect of DD coupling phase on CE performance.}
%	\label{NMSE_phase}
%\end{figure}

\begin{figure}[t]
	\centerline{\includegraphics[width=\columnwidth ]{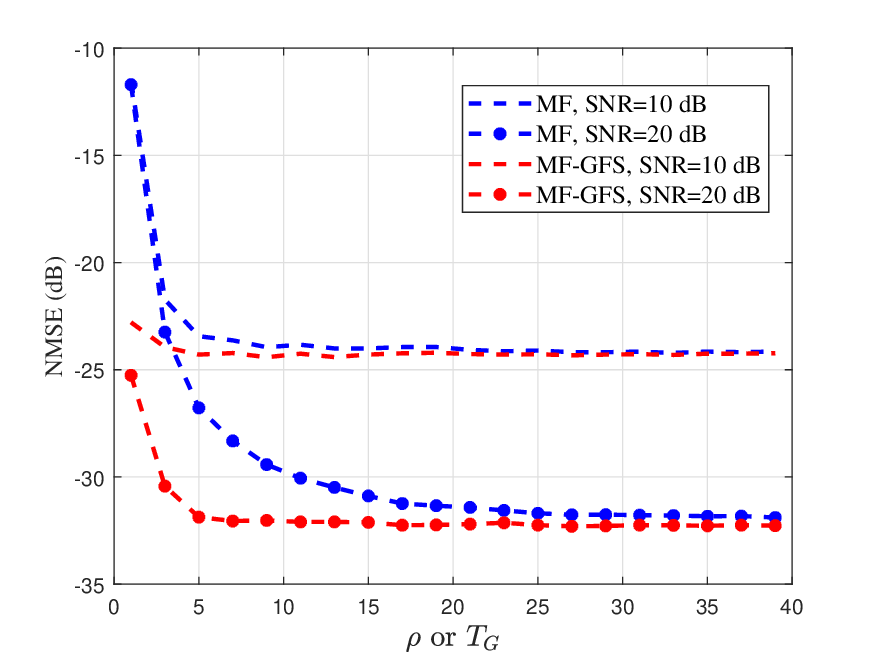}}
	\caption{Effects of $\rho$ on MF CE and $T_G$ on MF-GFS CE.}
	\label{MF_NMSE_rho1}
\end{figure}
%\begin{figure}[h]
%	\centerline{\includegraphics[width = 7cm ]{./fig/simulation/MF_NMSE_rho2.eps}}
%	\caption{NMSE of MF scheme at different $\rho$ vs.  SNR.}
%	\label{MF_NMSE_rho2}
%\end{figure}
\begin{figure}[t]
	\centerline{\includegraphics[width=\columnwidth ]{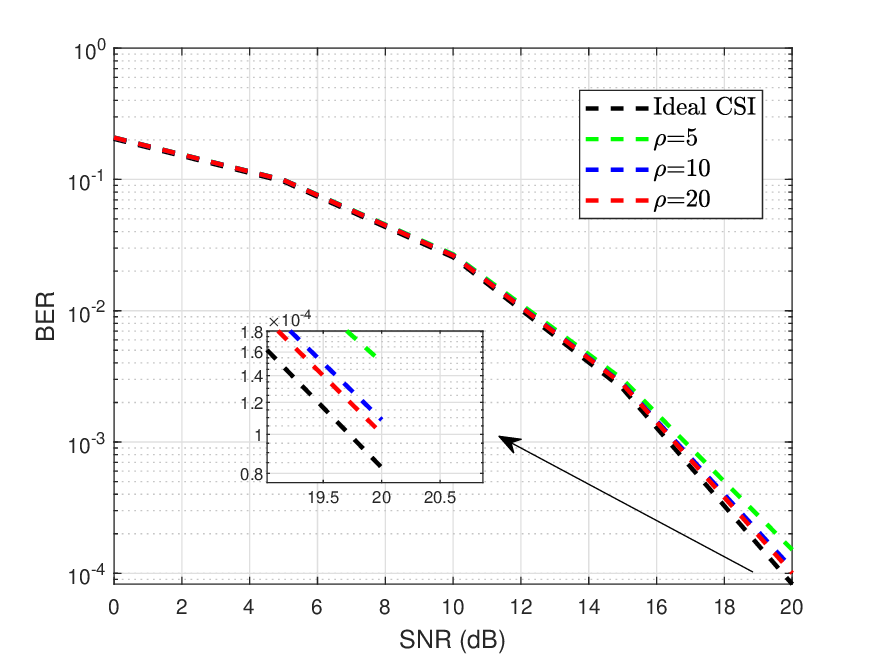}}
	\caption{BER comparison of the proposed MF scheme versus SNR for different $\rho$ values.}
	\label{MF_BER_rho}
\end{figure}

Fig. \ref{MF_NMSE_rho1} shows the effects of $\rho$ on MF and $T_G$ on MF-GFS CE for IDFD channel. Compared to the MF CE scheme, the MF-GFS CE scheme can achieve better performance with lower complexity.
Fig. \ref{MF_BER_rho} show the BER performance of the proposed MF scheme for IDFD channel with different SNR and $\rho$, respectively. The BER is obtained by LMMSE algorithm.
%As $\rho$ increases, both the NMSE and BER are improved.
%This performance improvement is due to the fact that as $\rho$ increases, the size of IU will decrease accordingly, which allows the fractional Doppler to be estimated more accurately.
As $\rho$ increases, the size of IU correspondingly decreases, enabling more accurate estimation of the fractional Doppler, which leads to improvements in both NMSE and BER.

%Fig. \ref{MF_NMSE_rho1} show the NMSE performance of the MF scheme at different $\rho$ for IDFD channels. 
%The NMSE performance of the MF PE scheme improves as the $\rho$ increases. 
%In addition, when $\rho$ reaches a certain value, approximately $20$ in this case, the NMSE performance no longer shows significant improvement. 
%Fig. \ref{MF_BER_rho} illustrates the BER performance of the MF scheme at different $\rho$ for IDFD channels. 
%It can be seen that the BER performance of the proposed MF scheme improves as the $\rho$ increases.

\begin{figure}[t]
	\centerline{\includegraphics[width=\columnwidth ]{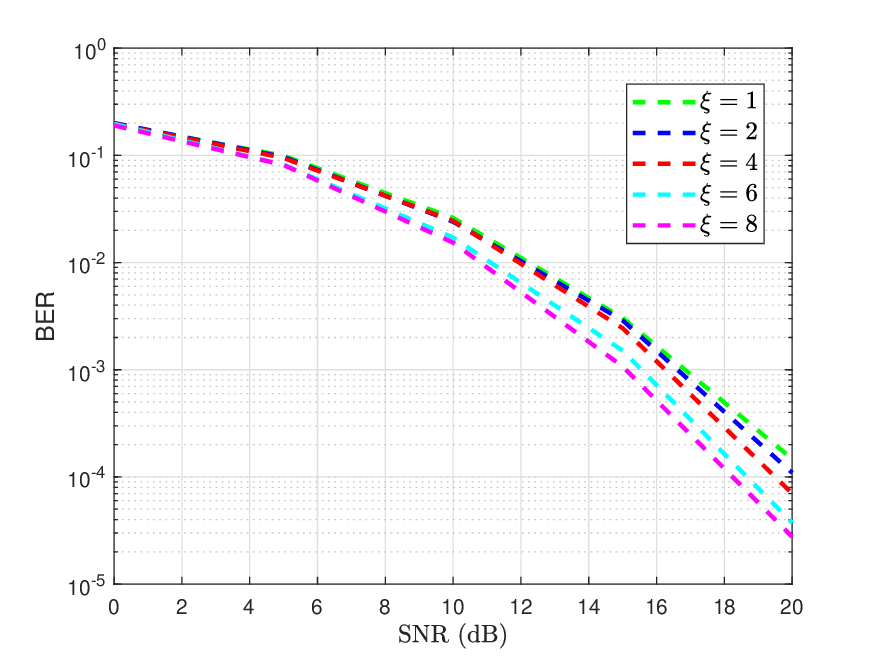}}
	\caption{BER comparison of the proposed MF scheme versus SNR for different $\xi$ values.}
	\label{MF_BER_xi}
\end{figure}
The impact of different $\xi$ on the BER of the MF scheme is shown in Fig. \ref{MF_BER_xi}. It can be observed that the BER improves as the $\xi$ increases.
In this case, the interference between different paths will decrease, making the estimated parameters more accurate, and hence improved BER performance.
In addition, according to Appendix B and Fig. \ref{IM_xi_N}, as $\xi$ increases, the orthogonality performance of $\boldsymbol{A}_{\mathrm{T}}\left( \boldsymbol{l}_{\tau},\boldsymbol{k}_{\nu} \right) $ is improved, thus leading to further improved CE and BER performances.
%the improved orthogonality for $\boldsymbol{A}_{\mathrm{T}}\left( \boldsymbol{l}_{\tau},\boldsymbol{k}_{\nu} \right) $ leads to a reduction in the error of $\boldsymbol{A}_{\mathrm{T}}^{\mathrm{H}}\left( \boldsymbol{l}_{\tau},\boldsymbol{k}_{\nu} \right) \boldsymbol{A}_{\mathrm{T}}\left( \boldsymbol{l}_{\tau},\boldsymbol{k}_{\nu} \right) $ approximating the identity matrix $\boldsymbol{I}$, which further improves the PE performance and BER performance.

\begin{figure}[t]
	\centerline{\includegraphics[width=\columnwidth ]{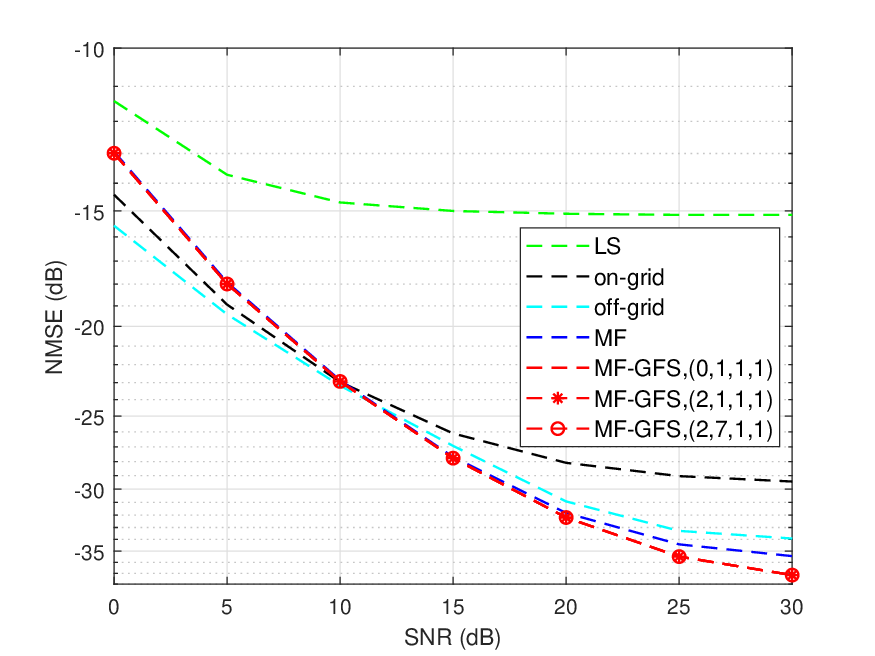}}
	\caption{Comparison of NMSE performance.}
	\label{NMSE_all}
\end{figure}
\begin{figure}[t]
	\centerline{\includegraphics[width=\columnwidth ]{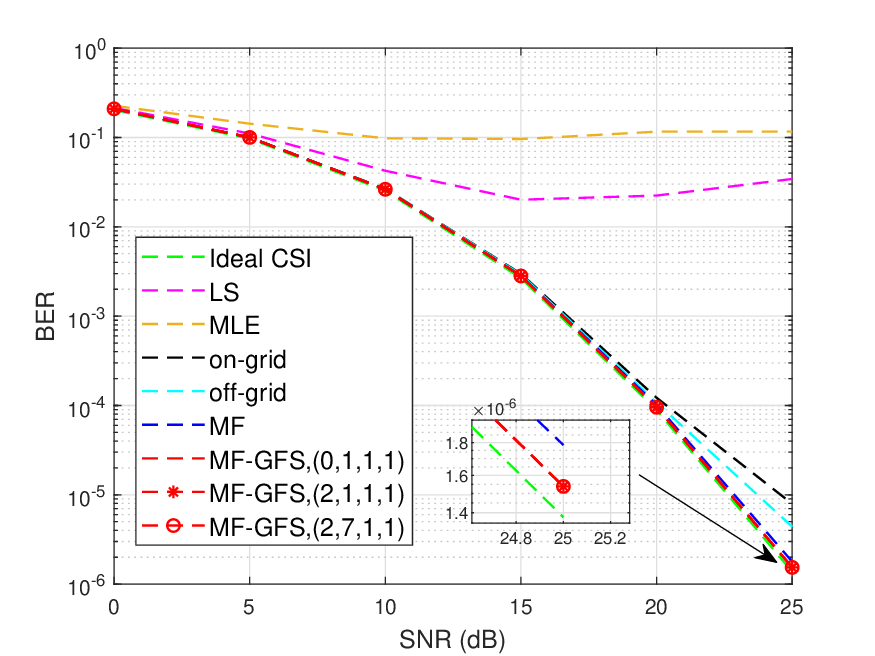}}
	\caption{Comparison of BER performance.}
	\label{BER_all}
\end{figure}
\begin{figure}[t]
	\centerline{\includegraphics[width=\columnwidth ]{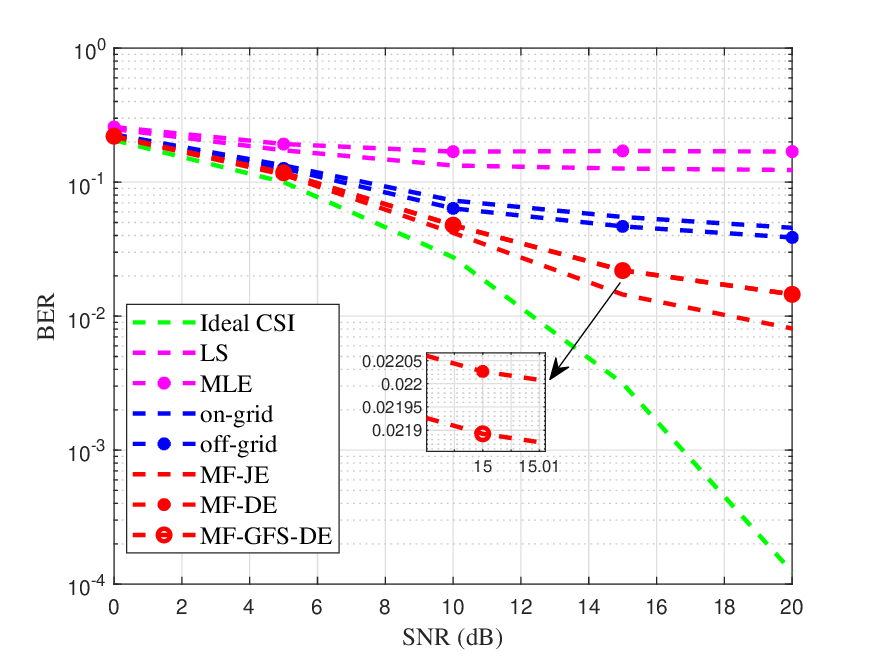}}
	\caption{Comparison of BER performance for FDFD channels.}
	\label{BER_all_FDD}
\end{figure}
The NMSE and BER performances of different schemes for IDFD channels are compared in Fig. \ref{NMSE_all} and Fig. \ref{BER_all}, respectively. We set $\rho=15$ and $T_G=8$. In addition, the performance of the MF-GFS scheme with different GFN, distinguished by $\left(a,b,p,q\right)$, is also compared.
%For the measurement matrix construction in the OMP scheme, the virtual sampling of a Doppler grid is 0.5.
%For simplicity, the mark ‘+’ corresponding to the solid line indicates that the $\hat{h}_w\left( m,n \right)$ in \eqref{NMSE} or the LMMSE detection algorithm is considered without accounting for the DD coupling phase $e^{\small{\frac{-j2\pi}{N}k_{\nu _i}l_{\tau _i}}}$.
%For simplicity, the mark \text{\ding{73}} denotes the AFDM framework in \cite{Bemani_3}.
The performance of the proposed MF scheme and MF-GFS scheme are significantly improved compared to that of other schemes.
%The LS and MLE schemes fail to accurately estimate the fractional Doppler, which results in its performance degradation.
Since the size of the IU estimated by the MF-GFS scheme is significantly smaller than that of MF, it can be seen that the performance of the proposed MF-GFS scheme outperforms that of the MF scheme at high SNR. In addition, since the ratio of two consecutive GFNs is generally close to the golden ratio, different GFNs yield similar IU divisions, thereby resulting in a negligible impact on the performance of the MF-GFS scheme.

Fig. \ref{BER_all_FDD} is a comparison of BER performance for FDFD channels.
As analyzed in subsection \Rmnum{3}-B-3), local peaks induced by $\iota _{\tau _i}$ exacerbate ICCI and inter-path interference, leading to deteriorated CE and BER performances. However, the proposed MF CE scheme outperforms existing schemes.

\section{Conclusions}
This paper has studied an MF-based CE scheme for AFDM systems, by upholding the key idea of sequentially estimating the parameters of each path. After analyzing the channel in the DAFT domain, we have proposed two low-complexity methods to construct the channel matrix by exploiting the inherent structural properties of the channel matrix, reducing complexity by at least two orders of magnitude.
We have proposed the MF-JE CE scheme, which decouples multipath estimation by exploiting path orthogonality and separability.
Furthermore, to further reduce the redundant computation in estimating fractional parameters, the MF-GFS-DE CE scheme was developed. This scheme incorporated the GFS algorithm and decoupled joint DD estimation at the cost of a slight performance loss.
Both schemes are extended to the typical broadband systems. The simulation results have shown that the proposed schemes offer more significant advantages over existing CE schemes in terms of improved communication performances and lower complexities.

% 附录有多个section时
\appendices
\section{Proof of Theorem \ref{thm1}}
To analyze the impact of each path, a more detailed I/O relationship is required. 
Considering the noise-free version of \eqref{r_t}, its corresponding discrete-time signal expression is
\begin{equation}\label{r_n_expanded}
	\begin{aligned}
		r\left[ \bar{n} \right] =\frac{1}{\sqrt{N}}\sum_{i=0}^{P-1}{h_i\sum_{n=0}^{N-1}{x\left[ n \right]}}e^{\frac{j2\pi}{N}k_{\nu _i}\left( \bar{n}-l_{\tau _i} \right)} \hspace{40pt} \\ \times e^{j2\pi \left( c_2n^2+c_1\left( \bar{n}-l_{\tau _i} \right) ^2+\frac{n}{N}\left( \bar{n}-l_{\tau _i} \right) +d_{\bar{n},n}\iota _{\tau _i} \right)},
	\end{aligned}
\end{equation}
where $d_{\bar{n},n}=\sum_{q=0}^{2Nc_1}{\beta _{\bar{n},n}\left( q \right)}$, $\beta _{\bar{n},n}\left( q \right) $ is given by
\begin{equation}
	\beta _{\bar{n},n}\left( q \right) =\begin{cases}	q,&		\lfloor \frac{t_{n,q}}{T_s} \rfloor +1\le \bar{n}\le \lfloor \frac{t_{n,q+1}}{T_s} \rfloor\\	0,&		\mathrm{otherwise}\\\end{cases}.
\end{equation}
Note that once $N$ and $c_1$ are fixed, $d_{\bar{n},n}$ is determined and can thus be pre-calculated.

Substituting \eqref{r_n_expanded} and the DAFT kernel $\phi _m\left( t \right) $ into \eqref{y_m}, The noise-free version of $y\left[ m \right]$ can be expressed as \eqref{DD-1}.
%Substituting \eqref{AFDM modulation} and \eqref{r_n} into \eqref{y_m}, we have \eqref{y_m_IO}, as shown at the top of this page.
%Substituting the kernel $F_{c_1,c_2}\left( \cdot \right) $ of DAFT into \eqref{y_m_IO} and further expanding it, the received signal can be further expressed as \eqref{y_m_I0_2}, as shown at the top of this page.
%Combining the terms containing $\bar{n}$ together, $y\left( m \right) $ can be expressed as
%By grouping the terms involving $\bar{n}$ together, $y\left( m \right) $ can be expressed as
%\begin{equation}
%	y(m)=\sum_{n=0}^{N-1}{x\left( n \right) h_w\left( m,n \right)}+w\left( n \right) ,
%\end{equation}
%where $h_w\left( m,n\right) $ is given by \eqref{h_w} at the top of this page.
%%Based on the Euler's formula, \eqref{F} can be rewritten as \eqref{F_2}, as shown at the top of this page.
%Therefore, the proof is complete.
Finally, the square bracket in \eqref{DD-1} is replaced by $H_\mathrm{eff}\left[ m,n\right] $, which completes the proof.

\begin{figure*}[t] % hb底部，ht为头部
	\centering % 公式居中
	\hrulefill % 添加一条水平线
	%	\vspace*{8pt} % 调整线与公式之间的距离
	\begin{align}
		y\left[ m \right] &=\sum_{n=0}^{N-1}{x\left[ n \right] \left[ \sum_{i=1}^P{h_i\underset{\alpha \left( m,n,l_{\tau _i},k_{\nu _i} \right)}{\underbrace{e^{\small{\frac{-j2\pi}{N}}\left( -Nc_1l_{\tau _i}^{2}+\left( n+k_{\nu _i} \right) l_{\tau _i}+Nc_2\left( m^2-n^2 \right) \right)}}}\underset{\mathcal{F} \left( m,n,l_{\tau _i},k_{\nu _i} \right)}{\underbrace{\frac{1}{N}\sum_{\bar{n}=0}^{N-1}{e^{\small{\frac{-j2\pi}{N}}\left( m-n+2Nc_1l_{\tau _i}-k_{\nu _i} \right) \bar{n}-Nd_{\bar{n},n}\iota _{\tau _i}}}}}} \right]} ,\label{DD-1}
		\\ &=\sum_{n=0}^{N-1}{x\left[ n \right] H_\mathrm{eff}\left[ m,n \right]}+w\left[ n \right]. 
	\end{align}

\end{figure*}

%\begin{figure*}[t] % hb底部，ht为头部
%	\centering % 公式居中
%	\hrulefill % 添加一条水平线
%	%	\vspace*{8pt} % 调整线与公式之间的距离
%	\begin{equation} \label{h_w}
%		\begin{aligned}
%			h_w\left( m,n \right) =&\frac{1}{N}\sum_{i=1}^P{h_ie^{\small{\frac{-j2\pi}{N}}\left( -Nc_1l_{\tau _i}^{2}+\left( n+k_{\nu _i} \right) l_{\tau _i}+Nc_2\left( m^2-n^2 \right) \right)}\sum_{\bar{n}=0}^{N-1}{e^{\small{\frac{-j2\pi}{N}}\left( m-n+2Nc_1l_{\tau _i}-k_{\nu _i} \right) \bar{n}-Nd_{\bar{n},n}}}}\\=&\sum_{i=1}^P{h_i\alpha \left( m,n,l_{\tau _i},k_{\nu _i} \right) \mathcal{F} \left( m,n,l_{\tau _i},k_{\nu _i} \right)}.
%		\end{aligned}
%	\end{equation}
%\end{figure*}

\section{Proof of Theorem \ref{thm2}}
\begin{figure*}[t] % hb底部，ht为头部
	\centering % 公式居中
	\hrulefill % 添加一条水平线
	%	\vspace*{8pt} % 调整线与公式之间的距离
	%\begin{equation}
	\begin{align}
		\left| \boldsymbol{a}_{i}^{\mathrm{H}}\boldsymbol{a}_j \right|&=\left| \sum_{m=0}^{N-1}{\frac{1}{N}\sum_{n_i=0}^{N-1}{e^{\small{\frac{j2\pi}{N}}\left( (m-n_p+2Nc_1l_{\tau _i}-k_{\nu _i})n_i-Nd_{n_i,n_p}\iota _{\tau _i}\label{AA-1} \right)}\frac{1}{N}\sum_{n_j=0}^{N-1}{e^{\small{\frac{-j2\pi}{N}}\left( (m-n_p+2Nc_1l_{\tau _j}-k_{\nu _j})n_j-Nd_{n_j,n_p}\iota _{\tau _j} \right)}}}} \right|
		\\&=\frac{1}{N}\left| \sum_{n_i=0}^{N-1}{\sum_{n_j=0}^{N-1}{e^{\small{\frac{-j2\pi}{N}\left[ -n_p\left( n_j-n_i \right) +2Nc_1(l_{\tau _j}n_j-l_{\tau _i}n_i)-\left( k_{\nu _j}n_j-k_{\nu _i}n_i \right) -N\left( d_{n_j,n_p}\iota _{\tau _j}-d_{n_i,n_p}\iota _{\tau _i} \right) \right]}}\underset{\delta \left( n_j-n_i \right)}{\underbrace{\frac{1}{N}\sum_{m=0}^{N-1}{e^{\small{\frac{-j2\pi}{N}}m\left( n_j-n_i \right)}}}}}} \right|.\label{AA-2}
	\end{align}
	%\end{equation}
\end{figure*}
%\begin{figure*}[t] % hb底部，ht为头部
%	\centering % 公式居中
%	\hrulefill % 添加一条水平线
%	%	\vspace*{8pt} % 调整线与公式之间的距离
%	%\begin{equation}
%	\begin{align}
%		\left| \boldsymbol{a}_{\mathrm{T}}^{\mathrm{H}}\left( l_{\tau _1},k_{\nu _1} \right) \boldsymbol{a}_{\mathrm{T}}\left( l_{\tau _2},k_{\nu _2} \right) \right|
%		\approx& \left|\mathrm{sinc}\left\{ 2Nc_1(l_{\tau _2}-l_{\tau _1})-\left( k_{\nu _2}-k_{\nu _1} \right) \right\} \right|\label{AA-7}\\
%		\approx& \left|\mathrm{sinc}\left\{ \left( 2\left( \small{\frac{\nu _{\max}}{\varDelta f}}+\xi \right) +1 \right) (\tau _2-\tau _1)N\varDelta f-\left( \nu _2-\nu _1 \right) \frac{1}{\varDelta f} \right\} \right|\label{AA-8}.
%	\end{align}
%	%\end{equation}
%\end{figure*}
%For convenience, the two columns in $\boldsymbol{A}_{\mathrm{T}}\left( \boldsymbol{l}_{\tau},\boldsymbol{k}_{\nu} \right) $ are denoted as $\boldsymbol{a}_{\mathrm{T}}\left( l_{\tau _1},k_{\nu _1} \right) $ and $\boldsymbol{a}_{\mathrm{T}}\left( l_{\tau _2},k_{\nu _2} \right) $, respectively.

\begin{figure}[t]
	\centerline{\includegraphics[width=\columnwidth ]{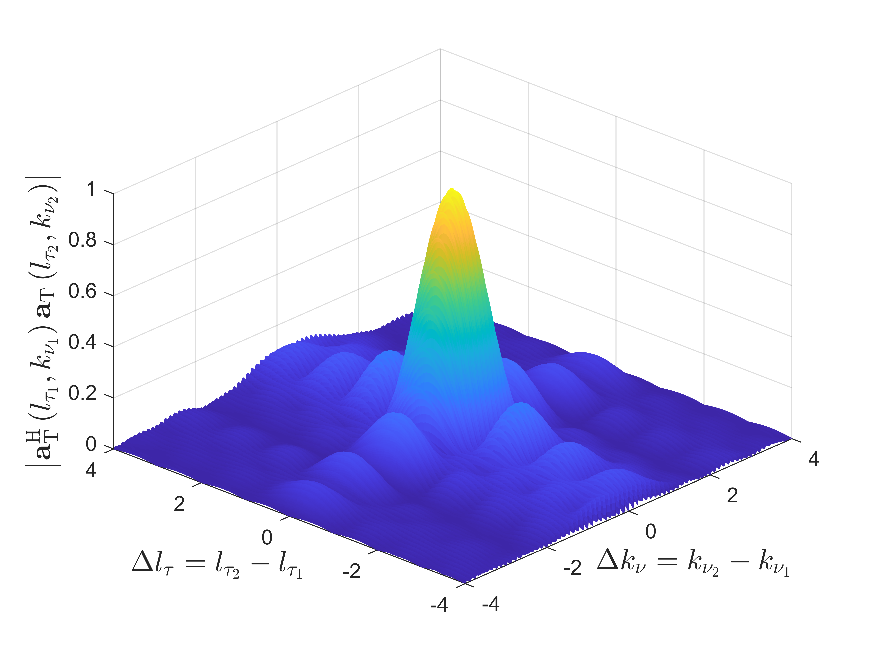}}
	\caption{Amplitude of $\left| \boldsymbol{a}_{\mathrm{T}}^{\mathrm{H}}\left( l_{\tau _i},k_{\nu _i} \right) \boldsymbol{a}_{\mathrm{T}}\left( l_{\tau _j},k_{\nu _j} \right) \right|$.}
	\label{Orthogonality_of_A}
\end{figure}

	For simplicity, we use $\left| \boldsymbol{a}_{i}^{\mathrm{H}}\boldsymbol{a}_j \right|$ to represent $\left| \boldsymbol{a}_{\mathrm{T}}^{\mathrm{H}}\left( l_{\tau _i},k_{\nu _i} \right) \boldsymbol{a}_{\mathrm{T}}\left( l_{\tau _j},k_{\nu _j} \right) \right|$.
$\left| \boldsymbol{a}_{i}^{\mathrm{H}}\boldsymbol{a}_j \right|$ can be expressed as \eqref{AA-1}-\eqref{AA-2}, as shown at the top of this page. Let $\varDelta l_{\tau}=l_{\tau _j}-l_{\tau _i}$, $\varDelta k_{\nu}=k_{\nu _j}-k_{\nu _i}$, $\left| \boldsymbol{a}_{i}^{\mathrm{H}}\boldsymbol{a}_j \right|$ can be represented as
\begin{equation} \label{AA-3}
	\begin{split}
	 \left| \boldsymbol{a}_{i}^{\mathrm{H}}\boldsymbol{a}_j \right|&=\frac{1}{N}\left| \sum_{\bar{n}=0}^{N-1}{e^{\small{\frac{-j2\pi}{N}\left\{ \left[ 2Nc_1\varDelta l_{\tau}-\varDelta k_{\nu} \right] \bar{n}-Nd_{\bar{n},n_p}\left( \iota _{\tau _j}-\iota _{\tau _i} \right) \right\}}}} \right|
	 \\&=\begin{cases}	1,&		\varDelta l_{\tau}=0,\varDelta k_{\nu}=0\\	\left| \sin\mathrm{c}\left( \varDelta k_{\nu} \right) \right|,&		\varDelta l_{\tau}=0,\varDelta k_{\nu}\ne 0\\	\frac{1}{N}\left| \sum_{\bar{n}=0}^{N-1}{\eta _1\eta _2} \right|,&		\varDelta l_{\tau}\ne 0,\varDelta k_{\nu}=0\\	\approx 0,&		\varDelta l_{\tau}\ne 0,\varDelta k_{\nu}\ne 0\\\end{cases},
	\end{split}
\end{equation} 
where $\eta _1=e^{\small{\frac{-j2\pi}{N}2Nc_1\varDelta l_{\tau}\bar{n}}}$, $\eta _2=e^{\small{j2\pi d_{\bar{n},n_p}\left( \iota _{\tau _j}-\iota _{\tau _i} \right)}}$.
Similar to the analysis in \eqref{F_FDFD} and \eqref{main_peak}, when $\varDelta l_{\tau}$ or $\varDelta k_{\nu}$ is zero, \eqref{AA-3} simplifies to a sinc function, peaking at $\varDelta k_{\nu}=0$ or $\varDelta l_{\tau}=0$, respectively. Therefore, when $\varDelta l_{\tau}=\varDelta k_{\nu}=0$ (i.e., $i= j$), the peak of $\left| \boldsymbol{a}_{\mathrm{T}}^{\mathrm{H}}\left( l_{\tau _i},k_{\nu _i} \right) \boldsymbol{a}_{\mathrm{T}}\left( l_{\tau _j},k_{\nu _j} \right) \right|$ is $1$. As $\varDelta l_{\tau}$ or $\varDelta k_{\nu}$ increases, $\left| \boldsymbol{a}_{\mathrm{T}}^{\mathrm{H}}\left( l_{\tau _i},k_{\nu _i} \right) \boldsymbol{a}_{\mathrm{T}}\left( l_{\tau _j},k_{\nu _j} \right) \right|$ is close to zero. This can also be seen in Fig. \ref{Orthogonality_of_A}. Therefore, for any $i\ne j$, $\left| \boldsymbol{a}_{\mathrm{T}}^{\mathrm{H}}\left( l_{\tau _i},k_{\nu _i} \right) \boldsymbol{a}_{\mathrm{T}}\left( l_{\tau _j},k_{\nu _j} \right) \right|$ generally satisfies
\begin{equation}
	\left| \boldsymbol{a}_{\mathrm{T}}^{\mathrm{H}}\left( l_{\tau _i},k_{\nu _i} \right) \boldsymbol{a}_{\mathrm{T}}\left( l_{\tau _j},k_{\nu _j} \right) \right|\leqslant \epsilon.
\end{equation}
In general, $\epsilon =\underset{i\ne j}{\max}\left| \boldsymbol{a}_{\mathrm{T}}^{\mathrm{H}}\left( l_{\tau _i},k_{\nu _i} \right) \boldsymbol{a}_{\mathrm{T}}\left( l_{\tau _j},k_{\nu _j} \right) \right| \ll 1$.
Therefore, this proof is completed.

\section{Proof of Theorem \ref{thm3}}
The proof of Theorem \ref{thm3} can be seen in \eqref{BB-1}-\eqref{BB-2}.
substituting $\boldsymbol{y}_T=\sum_{j=1}^P{x_ph_i\boldsymbol{a}_{\mathrm{T}}\left( l_{\tau _j},k_{\nu _j} \right)}$ and considering the orthogonality of $\boldsymbol{a}_{\mathrm{T}}\left( l_{\tau _i},k_{\nu _i} \right) $, \eqref{BB-1} is obtained.
%Substituting $\boldsymbol{y}_T=\sum_{j=1}^P{x_ph_i\boldsymbol{a}_{\mathrm{T}}\left( l_{\tau _j},k_{\nu _j} \right)}$ into $\left| \boldsymbol{\mathcal{T}} ^{\mathrm{H}}\left( \hat{\chi}_{i}^{\mathrm{I}}+\chi ^f \right) \boldsymbol{y}_{\mathrm{T}} \right|^2$ yields \eqref{BB-1}. From the proof of Theorem \ref{thm2}, it can be observed that vectors $\boldsymbol{a}_{\mathrm{T}}\left( l_{\tau _i},k_{\nu _i} \right) $ corresponding to different paths are approximately orthogonal. In addition, $\boldsymbol{a}_{\mathrm{T}}\left( l_{\tau _i},k_{\nu _i} \right) $ also has the same structure as $\boldsymbol{\mathcal{T}} ^{\mathrm{H}}\left( \hat{\chi}_{i}^{\mathrm{I}}+\chi ^f \right) $. Therefore, for \eqref{BB-2}, only $\boldsymbol{\mathcal{T}} ^{\mathrm{H}}\left( \hat{\chi}_{i}^{\mathrm{I}}+\chi ^f \right) \boldsymbol{a}_{\mathrm{T}}\left( l_{\tau _i},k_{\nu _i} \right) $ exists.
From \eqref{AA-3}, with $\varDelta l_{\tau}=l_{\tau _i}-\hat{l}_{\tau _i}\approx 0$ and $\bar{k}_{\nu _i}\approx\hat{\bar{k}}_{\nu _i}$, we obtain \eqref{BB-2}.
In addition, $\kappa _{\nu}-\kappa _{\nu _i}\in \left[ -1,1 \right] $.
Therefore, $\left| \boldsymbol{a}_{\mathrm{T}}^{\mathrm{H}}\left( \hat{\bar{l}}_{\tau _i}+\hat{\iota}_{\tau _i},\hat{\bar{k}}_{\nu _i}+\kappa _{\nu} \right) \boldsymbol{y}_T \right|^2$ is a bounded unimodal function, with its extremum achieved at $\kappa _{\nu}=\kappa _{\nu _i}$. Therefore, this proof is completed.

\begin{align}
	&\left| \boldsymbol{a}_{\mathrm{T}}^{\mathrm{H}}\left( \hat{\bar{l}}_{\tau _i}+\hat{\iota}_{\tau _i},\hat{\bar{k}}_{\nu _i}+\kappa _{\nu} \right) \boldsymbol{y}_T \right|^2 \notag
	\\&\hspace{30pt}\approx \left| x_ph_i\boldsymbol{a}_{\mathrm{T}}^{\mathrm{H}}\left( \hat{l}_{\tau _i},\hat{\bar{k}}_{\nu _i}+\kappa _{\nu} \right) \boldsymbol{a}_{\mathrm{T}}\left( l_{\tau _i},k_{\nu _i} \right) \right| \label{BB-1}
	\\&\hspace{30pt}\approx \left| x_ph_i \right|\mathrm{sinc}\left( \kappa _{\nu}-\kappa _{\nu _i} \right). \label{BB-2}
\end{align}

\newpage

% your own custom macro containing the $\backslash${\tt{includegraphics}} command to make things
 
\vspace{11pt}
%
%\bf{If you include a photo:}\vspace{-33pt}
%\begin{IEEEbiography}[{\includegraphics[width=1in,height=1.25in,clip,keepaspectratio]{fig1}}]{Michael Shell}
%Use $\backslash${\tt{begin\{IEEEbiography\}}} and then for the 1st argument use $\backslash${\tt{includegraphics}} to declare and link the author photo.
%Use the author name as the 3rd argument followed by the biography text.
%\end{IEEEbiography}

\vfill

\end{document}